\documentclass{aa}  
\usepackage{color}
\usepackage{graphicx}
\usepackage{placeins}
\usepackage{txfonts}
\usepackage[]{hyperref} % To add links in your PDF file
\usepackage{booktabs} % for much better looking tables
\usepackage{array} % for better arrays (eg matrices) in maths
\usepackage{paralist} % very flexible & customisable lists (eg. enumerate/itemize, etc.)
\usepackage{verbatim} % adds environment for commenting out blocks of text & for better verbatim
\usepackage{ulem}
\usepackage{xspace}
\usepackage{siunitx}
\usepackage{amssymb}
\usepackage{capt-of}
\usepackage{amsmath}
\usepackage{amssymb}
\usepackage{natbib}
\bibpunct{(}{)}{;}{a}{}{,} % to follow the A&A style

\graphicspath{{./img/}}

\makeatletter
\renewcommand*\aa@pageof{, page \thepage{} of \pageref*{LastPage}}
\makeatother

\addto\extrasenglish{%
}

\begin{document} 

    \title{CRIRES$^+$ transmission spectroscopy of WASP-127\,b}\subtitle{Detection of the resolved signatures of a supersonic equatorial jet and cool poles in a hot planet}

   \author{L. Nortmann \inst{1} \and
            F. Lesjak \inst{1} \and 
            F. Yan \inst{2} \and
            D. Cont \inst{3,4}  \and
            S. Czesla\inst{5} \and
            A. Lavail\inst{6} \and      
            A. D. Rains\inst{7} \and
            E. Nagel\inst{1} \and
            L. Boldt-Christmas\inst{7}\and
            A.~Hatzes\inst{5} \and           
            A. Reiners\inst{1} \and
            N. Piskunov\inst{7} \and
            O. Kochukhov\inst{7} \and
            U. Heiter\inst{7} \and
            D. Shulyak \inst{8}\and
            M. Rengel\inst{9} \and
            U. Seemann\inst{10} }

   \institute{
            Institut f\"ur Astrophysik und Geophysik, Georg-August-Universit\"at, Friedrich-Hund-Platz 1, 37077 G\"ottingen, Germany \\
            email: Lisa.Nortmann@uni-goettingen.de\and
            Department of Astronomy, University of Science and Technology of China, Hefei 230026, China \and
            Universitäts-Sternwarte, Fakultät für Physik, Ludwig-Maximilians-Universität München, Scheinerstr. 1, 81679 München, Germany \and
            Exzellenzcluster Origins, Boltzmannstraße 2, 85748 Garching, Germany \and
            Thüringer Landessternwarte Tautenburg, Sternwarte 5, 07778 Tautenburg, Germany \and
            Institut de Recherche en Astrophysique et Planétologie, Université de Toulouse, CNRS, IRAP/UMR 5277, 14 avenue Edouard Belin, F-31400, Toulouse, France \and
            Department of Physics and Astronomy, Uppsala University, Box 516, 75120 Uppsala, Sweden \and       
            Instituto de Astrofísica de Andalucía - CSIC, Glorieta de la Astronomía s/n, 18008 Granada, Spain\and
            Max-Planck-Institut für Sonnensystemforschung, Justus-von-Liebig-Weg 3, 37077 Göttingen, Germany \and
            European Southern Observatory, Karl-Schwarzschild-Str. 2, 85748 Garching bei München, Germany }

   \date{Received  19 April 2024; accepted 1 October 2024}
   
% \abstract{}{}{}{}{} 
% 5 {} token are mandatory

  \abstract 
{General circulation models of gas giant exoplanets predict equatorial jets that drive inhomogeneities in the atmospheric physical parameters across the planetary surface.}
{We studied the transmission spectrum of the hot Jupiter WASP-127\,b during one transit in the K band with CRIRES$^+$.}
{Telluric and stellar signals were removed from the data using \texttt{SYSREM} and the planetary signal was investigated using the cross-correlation technique. After detecting a spectral signal indicative of atmospheric inhomogeneities, we employed a Bayesian retrieval framework with a two-dimensional modelling approach tailored to address this scenario.}
{We detected strong signals of H$_2$O and CO, which exhibited not one but two distinct cross-correlation peaks. The double-peaked  signal can be explained by a supersonic equatorial jet and muted signals at the planetary poles, with the two peaks representing the signals from the planet's morning and evening terminators. We calculated an equatorial jet velocity of $7.7\pm0.2$ km~s$^{-1}$ from our retrieved overall equatorial velocity and the planet’s tidally locked rotation, and derive distinct atmospheric properties for the two terminators as well as the polar region. Our retrieval yields a solar C/O ratio and metallicity, and shows that the muted signals from the poles can be explained by either significantly lower temperatures or a high cloud deck. It provides tentative evidence for the morning terminator to be cooler than the evening terminator by $-175^{+133}_{-117}$ K.}
{Our detection of CO challenges previous non-detections of this species in WASP-127b’s atmosphere. The presence of a clear double-peaked signal highlights the importance of taking planetary three-dimensional structure into account during interpretation of atmospheric signals. The measured supersonic jet velocity and the lack of signal from the polar regions, representing a detection of latitudinal inhomogeneity in a spatially unresolved target, showcases the power of high-resolution transmission spectroscopy for the characterisation of global circulation in exoplanet atmospheres.}


   \maketitle
%
%-------------------------------------------------------------------
\section{Introduction}
While we can readily resolve and study the three-dimensional (3D) structure of Solar System planets \citep{1, 2}, similar investigations of exoplanets are challenging due to their vast distances from us. Nonetheless, the study of atmospheric inhomogeneities in these planets has started to make significant progress in recent years. Exoplanet atmospheres can be explored using spectrophotometric observations. These studies are particularly favourable for transiting planets, for which not only dayside emission and reflection spectra can be obtained by observing
phases close to the secondary eclipse but also transmission spectra can be obtained by
probing the planet terminator during a transit event. During the transit, part of the star light passes though the upper atmosphere of the planet and is imprinted with the unique pattern of absorption lines of atoms and molecules present in the planet atmosphere.

Previous studies have investigated atmospheric inhomogeneities by observing time resolved changes of the planetary spectra, and thus probing different longitudes of the planet as they rotate in or out of view \citep{3,4,5,6,7,8,9,10}. These studies have been conducted at both low and high spectral resolution. 
When observing at high resolution ($\mathcal{R}\sim$100\,000) the dynamics of the atmospheric material can be measured through Doppler shifts of the   signals caused by their relative motions towards the observer 
\citep{11, 3}. Using this benefit, high wind speeds from jet streams and dayside-to-nightside winds as well as temperature differences can be deduced from observed line broadening and asymmetries \citep{12,2021MNRAS.502.1456K}.
Previous studies have investigated signals stemming from different parts of the planet atmospheres, including their morning and evening terminators, which were observed superimposed on each other and had to be disentangled through comparison with models containing combined absorption features of different regions \citep{6, 2023AJ....165..242G, 2024A&A...687A..49M} and comparing differences between ingress and egress line shapes \citep{4,7}.
One advantage that was used to disentangle signals in the time domain was that very close-in planets with very short orbital periods ($\sim$ 1-2 days) undergo significant changes in projected viewing angle between transit ingress and egress \citep{6,3,4,10}. This effect, however,  becomes progressively less pronounced for planets at larger distances from their host stars (i.e. with longer orbital periods).

Here we studied the transmission spectrum of the  hot Jupiter exoplanet WASP-127\,b, which  orbits its G-type host star on a 4.18-day orbit \citep{22}. The planet has one of the lowest densities ever recorded for an exoplanet \citep[$\rho = 0.07 \pm 0.01 \rho_\mathrm{Jup}$;][]{22}, which, in combination with its favourable star-planet radius ratio, makes it a prime target for atmospheric studies. The first indication of a feature-rich transmission spectrum on this planet was obtained at low resolution with the Andalucia Faint Object Spectrograph and Camera (ALFOSC) mounted on the 2.5m Nordic Optical Telescope (NOT) at the Roque de los Muchachos Observatory (ORM)  \citep{47}. These findings were later confirmed with higher precision using the OSIRIS instrument at the 10m Gran Telescopio Canarias (GranTeCan) \citep{48}, showing not only sodium and potassium absorption, but also a tentative detection of lithium in the planet. However, follow-up studies of the planet at high resolution in the optical wavelength range only measured a weak signal for sodium with ESPRESSO at the 8m Very Large Telescope (VLT) \citep{23} and HARPS \citep{24}, the latter being compatible with a non-detection. Using the Rossiter-McLaughlin effect, these high-resolution observations of the planet during transit revealed that the planet is orbiting its host on a misaligned orbit \citep{23, 49}. The atmosphere and orbit were further constrained by successful eclipse measurements with \textit{Spitzer}, which determined the planet’s dayside temperature as $T \approx 1400$ K \citep[$1454^{+42}_{-43}$ K and $1373^{+40}_{-41}$ K at  \mbox {3.6\,$\mu$m} and \mbox{4.5\,$\mu$m}, respectively;][]{50}. 
 Low-resolution space-based spectroscopy obtained in the H band with the WFC3 instrument on the \textit{Hubble} Space telescope (HST) led to a detection of water in the planet’s transmission spectrum \citep{51, 21}. An atmospheric retrieval study combining the HST and \textit{Spitzer} transit data led to conflicting carbon-to-oxygen ratios (C/O) depending on whether chemical equilibrium or free chemistry assumptions were adopted \citep[leading to values of \mbox{C/O $\sim 0.8$} and \mbox{C/O $\sim 0.0$}, respectively;][]{21}. This degeneracy was seemingly  solved through recent high-resolution observations of this target over a wide wavelength range in the near-infrared ($\sim980$ - 2500\,nm) using the SPIRou spectrograph, which yielded a detection of water (H$_2$O) and hydroxyl (OH), but no carbon monoxide (CO). The non-detection of CO led to strong upper limits on the CO abundance ($\log_{10}(\mathrm{CO}) < -4.0$) and favoured a disequilibrium case with a low C/O ratio for this planet in the joint retrieval of SPIRou + HST + \textit{Spitzer} data \citep{36}. The H$_2$O and OH signals found in this high-resolution study were detected to be strongly blueshifted (by approximately 10\,km s$^{-1}$) from the planet's rest frame and the authors discussed the possibility of this signal being only part of a broadened velocity signature, with other parts of the signal hidden within the noise.

For this work we used the CRyogenic InfraRed Echelle Spectrograph (CRIRES$^+$) in the K band (1972 - 2452 nm) and independently investigated the atmospheric composition and its atmospheric dynamics at a spectral resolution of $\mathcal{R}\approx$140\,000. 
The instrument has recently been used to successfully  investigate exoplanet atmospheres,   in transmission and in emission spectroscopy \citep[e.g.][]{2022AJ....164...79H,32,29,2023MNRAS.525.2985R,2024A&A...688A.191G,2024MNRAS.531.2356P,2024A&A...688A.206C}.

We report the detection of both H$_2$O and CO using the cross-correlation method. We further report the detection of two peaks in the velocity profile for each of these species, located at approximately $\pm$ 8\,km s$^{-1}$. These peaks remained resolved throughout the entire duration of the planetary transit. We found that these two peaks can be explained as representing the signals stemming from the two limbs of the planet. We used these spectroscopically resolved signals to retrieve the atmospheric properties of the two equatorial regions of the morning limb and the evening limb, and for the poles, via a Bayesian retrieval framework. The retrieval   utilized a simplified two-dimensional (2D) atmospheric model, which incorporates the freedom of varying dynamics and atmospheric conditions at different latitudes and longitudes of the planet. This approach is analogous to previous studies using  2D retrievals to disentangle the components of the morning and evening terminator in unresolved signals \citep{6, 2023AJ....165..242G, 36,2024A&A...687A..49M}, but additionally includes the option of latitudinal variations, further pushing the exploration of inhomogeneities in exoplanet atmospheres.

Our work is structured as follows. The observations are described in Sect. \ref{sec:obs}. The data reduction and pre-processing, such as the removal of stellar and telluric signals, is described in Sect. \ref{sec:dataredproc}. The detection of atmospheric species is described in Sect. \ref{sec:detec};   details on the model calculation are given in Sect. \ref{sec:1dmodels} and the cross-correlation analysis is described in Sect. \ref{sec:cc}. The results of the cross-correlation analysis are given and discussed in Sects. \ref{sec:cc_results} and \ref{sec:cc_discussion}, respectively. In Sect. \ref{sec:doublepeak} we discuss the interpretation of the double-peaked signals. The set-up of the 2D atmospheric retrieval is described in Sect. \ref{sec:retrievalsetup}.  The results of the retrieval are provided and discussed in Sect. \ref{sec:retrieval_results_discussion}.  Finally, we give conclusions about our work in Sect. \ref{sec:conclusions}.

\begin{figure}
\centering
\includegraphics[width=\hsize]{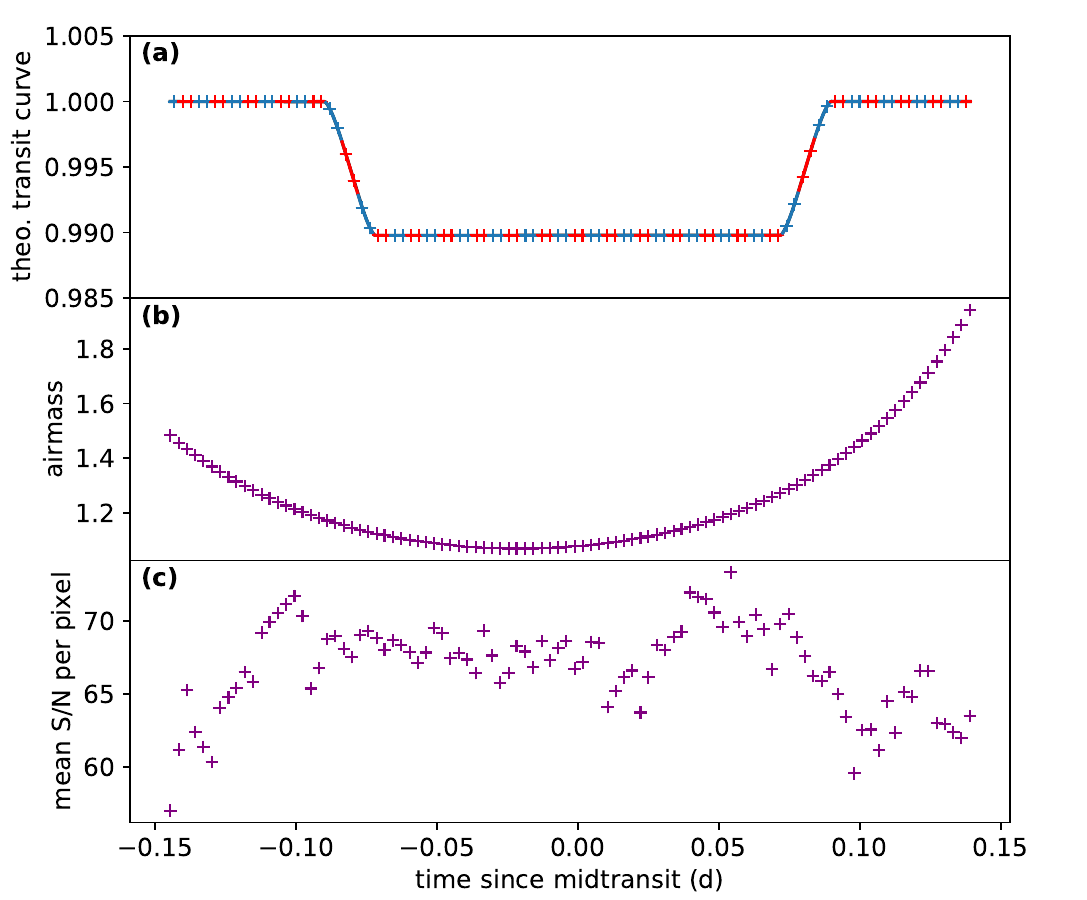}
    \caption{Transit coverage and observing conditions during the observation of WASP-127\,b on the night of 24--25 March 2022.  {Panel a}: Theoretical light curve based on literature parameters indicating the 98 exposures of 240s. Each data point corresponds to one exposure, and its width to the duration of the exposure. 
    Blue crosses indicate nodding position A and red crosses nodding position B. The transit including ingress and egress (i.e. the first to fourth contact T1-T4) is covered by spectra 20-81; the second to third contact (T2-T3) is covered by spectra 26-75. {Panel b}: 
    Progression of the airmass. {Panel c}:
    Mean signal-to-noise ratio (S/N) of the spectra over all covered wavelengths. }
     \label{fig:S1}
\end{figure}

\section{Observations}\label{sec:obs}
We observed the extrasolar planet WASP-127\,b during a transit event (i.e. when the planet passed in front of its host star in our line of sight) on the night 24-25 March 2022 in the K band using the upgraded infrared spectrograph CRIRES$^+$, mounted at the Nasmyth B focus of the 8m UT3 telescope at the VLT Facility of the European Southern Observatory (ESO) \citep{13}. The observation lasted for 6.6 hours (00:22 UT - 07:03 UT) and covered the whole transit event as well as a comparable amount of pre- and post-transit baseline (Fig. \ref{fig:S1} a).  The exposure time of 240\,s was purposefully chosen: short enough to avoid the planet signal spreading over several detector pixels, due to the Doppler shift caused by the planet's orbital motion \citep{2024A&A...683A.244B}. For comparison, the projected  orbital velocity of this planet changes by approximately 1 km s$^{-1}$ every 456\,s during transit.  In total we obtained 98 spectra, 62 of them in transit, including ingress and egress (i.e. between first and fourth contact, T1-T4), and  50 (of the 62) between the second and third contact, excluding ingress and egress (T2-T3) (see Fig. \ref{fig:S1} a). 
The airmass at the start of the observation was 1.5, reached a minimum of 1.07 during the transit, and then rose again to 1.9 towards the end of the observation in the post-transit coverage. The signal-to-noise ratio (S/N) per pixel fluctuated between 57-73. The progression of the airmass and mean S/N over the course of the observing night are shown in Fig. \ref{fig:S1} b and c, respectively. 
We used a slit width of 0.2'' and the K band setting K2148, which covers the wavelength range   1972 - 2452 nm. An ABBA nodding sequence with a separation of the two nodding positions of 6'' was employed to facilitate precise sky background subtraction. During the observations, the adaptive optics system was used and the guiding was done on the target itself.

\section{Data reduction and pre-processing\label{sec:dataredproc}}
\subsection{Extraction of 1D  spectra and wavelength solution}
The spectra were extracted from the 2D images using the CRIRES$^+$ Data Reduction Software (DRS) \texttt{CR2RES} (Version 1.1.4). Details can be found in Sect. \ref{App:extraction}.
During the observation the full width half maximum (FWHM) of the stellar point spread function was smaller than the slit width. This translated into a spectral resolution of $\mathcal{R}\approx$140\,000 and caused an offset between the wavelength solution of the science spectra and that of the calibration spectra (see Fig. \ref{fig:S2}). For this reason, instead of adopting the wavelength solution provided by the pipeline, which is based on the calibration spectra, we obtained a refined wavelength solution from the science frames themselves. This was done by fitting the position of the telluric lines with the ESO tool \texttt{molecfit} \citep{14} in two reference spectra taken at the two respective nodding positions. Further, we corrected for spectral drifts during the night by cross-correlating each individual spectrum with the reference spectra and using the measured offsets to  interpolate all spectra to the same wavelength grid. Details on the determination of the wavelength resolution and the wavelength offsets are given in Sect. \ref{App:wlsolution}. More details on the use of \texttt{molecfit} are described in Sect. \ref{App:molecfitccf}.

\subsection{Removal of stellar and telluric signals}
To investigate the planet spectrum, all stellar and telluric signals had to be removed. For this purpose all spectra were normalised to the same continuum level, and the deepest telluric lines, with flux below 20\% of the continuum level, were masked.
More details on the normalisation procedure can be found in Sect. \ref{app:normalization}. We then brought all spectra into the stellar rest frame, and removed stellar and telluric signals, which are quasi-static in time, using the detrending algorithm \texttt{SYSREM} \citep{15, 16}. 
We found that performing the detrending in stellar rest frame improved the removal of stellar residuals without compromising the removal of telluric residuals. This was tested by performing the analysis for both cases  (see Sect. \ref{app:alignmentofspectra} for more background information on the spectral alignment and its motivation). We ran \texttt{SYSREM} for 20 iterations and applied the correction in flux space following \citeauthor{17} \citeyear{17}, through division of the data by the summed up \texttt{SYSREM} models. The analysis described in the following sections was then performed on each of the 20 iterations.

\section{Detection of atmospheric species and velocity profile in the planetary transmission spectrum} \label{sec:detec}
After the correction of the quasi-static signals found by \texttt{SYSREM}, the data only contained noise and the planet signal, which is Doppler shifted by different amounts in each spectrum, due to the rapidly changing planetary velocity. To detect the planet signal in the noise, we used the 
high-resolution cross-correlation spectroscopy (HRCCS) method \citep{19,2012Natur.486..502B,2013A&A...554A..82D,16,2014ApJ...783L..29L}, meaning that we calculated the cross-correlation functions (CCFs) of the data and synthetic atmospheric models.

\subsection{Computation of synthetic 1D models for the cross-correlation}
\label{sec:1dmodels}
The models we used in this step of the analysis were calculated with \texttt{petitRADTRANS} \citep{20}. The input parameters for \texttt{petitRADTRANS} are a temperature-pressure profile ($T$-$p$ profile); the abundances of the molecules to be included in the model; and the system parameters, such as the planet radius $R_\mathrm{p}$, stellar radius $R_\mathrm{s}$, and planet surface gravity $\log_{10}(g_\mathrm{p})$. For our cross-correlation analysis we used models with solar C/O ratio and metallicity (i.e.  C/O = 0.55 and [Fe/H] = 0.0), and a $T$-$p$ profile based on the one retrieved for this planet from low-resolution spectroscopy by \cite{21} for equilibrium chemistry assumptions.

\subsubsection{$T$-$p$ profile}
We set up a pressure grid spanning from $10^{-6}$ to $10^2$ bar in 100 steps and calculated the temperatures for each pressure point using Eq. 29 in \cite{40}, which is implemented in \texttt{petitRADTRANS} as a function with five input parameters: the planetary internal temperature $T_\mathrm{int}$, the atmospheric equilibrium temperature $T_\mathrm{eq}$, the atmospheric opacity in the infrared (IR) wavelengths $\kappa_\mathrm{IR}$, the ratio of the optical to IR opacity $\gamma$, and the planet surface gravity $g_\mathrm{p}$.
For the planet surface gravity, we adopted the value of $\log_{10}(g_\mathrm{p} /($cm s$^{-2}))= 2.36$, which we calculated following \cite{41} using the literature system parameters from \cite{24}, which are summarised in Table \ref{tab:seidelparameters}. Our value for $\log_{10}(g_\mathrm{p})$ is consistent within the errors with the value given by \cite{22}. As \cite{21} use a slightly different parametrisation of the Guillot \mbox{$T$-$p$} profile, we cannot directly adopt all five parameters needed to replicate their $T$-$p$ profile from their work. However, we found that using our value for $g_\mathrm{p}$ and the values they provide for the opacities ($\log_{10}(\kappa_\mathrm{IR}) = -2.99$ and $\log_{10}(\gamma) = -1.92$), setting $T_\mathrm{int}= 500$ K and $T_\mathrm{eq} = 1135$ K in the \texttt{petitRADTRANS} parametrisation closely reproduces the $T$-$p$ profile shown in their manuscript \citep[][Fig. 11 top panel]{21}. 
\begin{table}[ht]\renewcommand{\arraystretch}{1.5}
 \caption[]{Parameters for the \mbox{WASP-127} system used in this manuscript.}\label{tab:seidelparameters}
\begin{tabular}{lll}
 \hline \hline
  Adopted parameter &
  Value
 \\ \hline
Planet radius    $R_\mathrm{p}$ ($R_\mathrm{Jup}$) &  $1.311^{+0.025}_{-0.029}$  \\
Planet mass   $M_\mathrm{p}$ ($M_\mathrm{Jup}$) &  $0.1647^{+0.0214}_{-0.0172}$  \\
Stellar radius $R_\mathrm{s}$ ($R_\odot$) & $1.33^{+0.025}_{-0.029}$ \\
Stellar mass $M_\mathrm{s}$ ($M_\odot$) &  $0.949^{+0.022}_{-0.019}$  \\
Orbital period    $P_\mathrm{orb}$ (days) & $4.17806203^{+0.00000088}_{-0.00000053}$\\
Orbital inclination $i$ ($^\circ$) & $87.84^{+0.36}_{-0.33}$\\
 $a/R_\mathrm{s}$& $7.81^{+0.11}_{-0.09}$\\%LEMN
Time of mid-transit $T_0$ (BJD$_\mathrm{TBD}$) & $2456776.62124^{+0.00023}_{-0.00028}$ \\
Systemic velocity $v_\mathrm{sys}$ (km\,s$^{-1}$) & -8.25 \\
Stellar radial velocity semi-& $0.022^{+0.003}_{-0.002}$& \\
~~~amplitude $K_\mathrm{s}$ (km\,s$^{-1}$) & 
 \\ \hline
\end{tabular}
\tablebib{The listed parameters are all from \citet{24}.}
\end{table}

\subsubsection{Atmospheric absorbers}
For the abundances we used the \texttt{poor\_mans\_nonequ\_chem} subpackage of \texttt{petitRADTRANS}, which  allowed us to interpolate pressure-dependent chemical abundances for each species based on the input $T$-$p$ profile,   C/O ratio, and metallicity [Fe/H].  This function uses a chemical grid calculated with \texttt{easyCHEM} \citep{42}.

We focused on the strongest absorbers expected in this wavelength range:  water (H$_2$O), carbon monoxide (CO), methane (CH$_4$), and carbon dioxide (CO$_2$). In the first step, we calculated separate spectra for each species, with each model only containing the absorption lines of one molecule.

Since we found no signals for CH$_4$ and CO$_2$, but significant CCF signals for both H$_2$O and CO, we then also calculated a model containing lines of both detected species simultaneously. 
A list of the used opacity information and the corresponding literature references can be found in Table \ref{tab:s2}.\\
Furthermore, we explored the option of the inclusion of an opacity deck as an additional source of absorption that could be caused by clouds or hazes. This grey opacity layer is referred to as a grey cloud deck in \texttt{petitRADTRANS}. When we refer to clouds in the remainder of the manuscript, we intend this grey opacity deck, which may have multiple potential origins. 
We find that a grey cloud layer at an atmospheric pressure of $\log_\mathrm{10}(P_\mathrm{c}/\mathrm{bar}) =-3$\,bar has an  effect on the shape of the model spectra comparable to that of any additional haze layer with the parameters proposed from retrievals of low-resolution data \citep{21}. We do not include a haze in our models  because the effects of both absorbers are interchangeable when only investigating a short wavelength range, as in our case. 
\begin{table}[ht]\renewcommand{\arraystretch}{1.5}
 \caption[]{Line list sources.}\label{tab:s2}
\begin{tabular}{l|c}
 \hline \hline
 Molecule&   Reference  \\ \hline
H$_2$O & \citep{52} \\
CO & \citep{52} \\
CH$_4$ & \citep{53} \\
CO$_2$ & \citep{52} \\
\hline
\end{tabular}
\end{table}

\subsubsection{Model resolution and normalisation}\label{sec:resolutionnorm}
Before we applied the cross-correlation of the models with the data, the models were normalised and brought to the spectral resolution provided by the instrument  for which we adopt a fixed value of $\mathcal{R}=140\,000$ as an approximation of the slightly variable resolution for each observed spectrum. This was done by convolution with a Gaussian kernel. For the convolution we  resampled the model onto a wavelength grid with equidistant sampling in velocity space. The sampling step was chosen so that we would maintain the same number of sampled points between the lowest and highest wavelength of the model.
The model normalisation was realised by dividing the model into small wavelength sections corresponding to the 18 wavelength sections covered by the data, and then dividing each of the 18 model sections by their respective mean value.

\subsection{Cross-correlation analysis}\label{sec:cc}
\subsubsection{Weighted cross-correlation}
We then performed a weighted cross-correlation by allowing velocity shifts $v$ from $-150$ km s$^{-1}$ to $+150$ km s$^{-1}$ in 1 km s$^{-1}$ steps. In each step, we calculated the weighted CCF as 
\begin{equation}CCF(v,t)=\sum_{j=0}^{N}\left(\frac{R_j(t) M_j (v)}{E_j(t)^2}\right)
,\end{equation}
where $R$ is the matrix of the residual spectra (i.e. data after \texttt{SYSREM} correction) with $t$ as the time index and $j$ as the pixel index, $M$ is the synthetic model shifted by the velocity $v$, and $E$ is the error corresponding to $R$. We performed the weighted cross-correlation for each of the 18 wavelength segments independently and then co-added the results of all segments. When summing over the segments we do not apply any additional weighting or exclusion of individual segments as is sometimes done by other groups \cite[e.g.][]{2021Natur.592..205G,2024A&A...686A.127B}. 
\begin{figure*}
\centering
\includegraphics[width=\hsize]{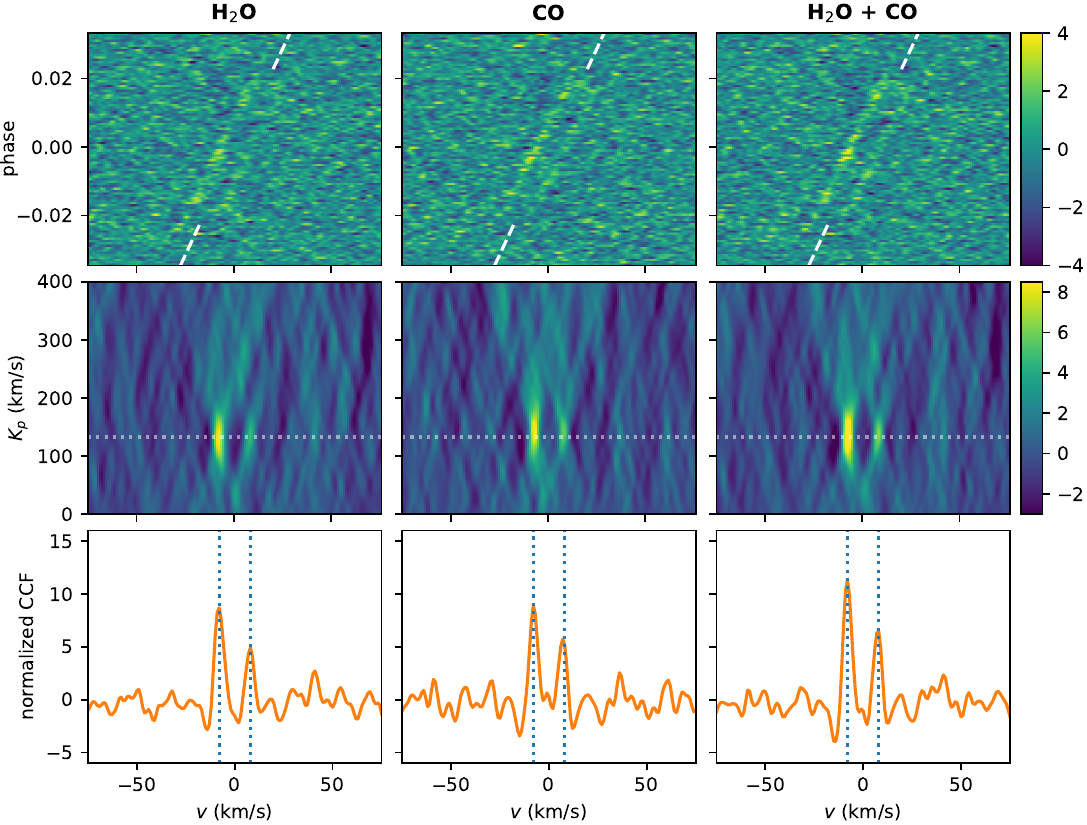}
  \caption{
  Results of the cross-correlation analysis of WASP-127\,b. The results are shown for cross-correlation with a pure H$_2$O model ({left column}), with a pure CO model ({middle column}), and  with a model containing both atmospheric species ({right column}). {Top panels:} Results for the CCF for each orbital phase. The radial velocity of the planetary orbital motion is indicated with a white dashed line. {Middle panels:} $K_\mathrm{p}$-$v$ map, i.e. the CCF co-added using different values for $K_\mathrm{p}$. The white dotted line indicates the value for $K_\mathrm{p}$ found in the literature. {Bottom panels:} Cut through the $K_\mathrm{p}$-$v$ map at the literature value for $K_\mathrm{p}$.}
     \label{fig:1}
\end{figure*}

\subsubsection{Calculation of $K_\mathrm{p}$-$v$ maps}
We then brought the CCF results into the planetary rest frame, assuming a circular planet orbit (i.e, $e = 0$) so that the planet velocity is given by $v_\mathrm{planet}=K_\mathrm{p} \sin(2 \pi \xi (t))$. Here  $\xi(t)$ denotes the orbital phase which we calculate using the ephemeris information from \cite[][summarised in Table \ref{tab:seidelparameters}]{24} and $K_\mathrm{p}$ is the planet's radial velocity semi-amplitude. While the value for $K_\mathrm{p}$ expected from literature parameters is $K_\mathrm{p, lit} = 133$ km s$^{-1}$, we repeated the calculation for $K_\mathrm{p}$ values between 0 and 400 km s$^{-1}$ in steps of 1 km s$^{-1}$ in order to be able to assess if the measured signals optimally align in the planet rest frame. In each step we aligned the spectra in the planet rest frame calculated using the presumed value for $K_\mathrm{p}$ and then co-added the CCFs of all spectra taken fully in-transit (i.e, between T2 and T3 %= spectra 26 to 75 
see Fig. \ref{fig:S1} a).

The results of the summed CCFs for each $K_\mathrm{p}$ value can be displayed in what is commonly referred to as a $K_\mathrm{p}$-$v$ map; this means that  the results are shown as a function of $K_\mathrm{p}$ and a velocity offset $v$. In our case, $v = 0$ corresponds to the planet’s rest velocity. Offsets of the signals from $v = 0$ may stem from uncertainties in the transit timing, from atmospheric dynamics (i.e. winds in the planetary atmosphere) or from uncertainties in the systemic velocity (here assumed to be $v_\mathrm{sys} = -8.25 \pm 0.89$~km~s$^{-1}$ based on the value from the Gaia catalogue; \citeauthor{43} \citeyear{43}).

To evaluate the significance of any signal, we normalised the entire $K_\mathrm{p}$-$v$ map by the standard deviation of values outside the region containing the signal. 
Here we used the two regions encompassed between \mbox{$v$ = [ -75 km s$^{-1}$, -20 km s$^{-1}$]} and \mbox{$v$ = [+20 km s$^{-1}$, +75 km s$^{-1}$]} over the entire $K_\mathrm{p}$ range (\mbox{i.e. [0 km s$^{-1}$, 400 km s$^{-1}$]}) to calculate the standard deviation representative of the noise of the $K_\mathrm{p}$-$v$ map outside the region containing peak signal. We then divided all values of the $K_\mathrm{p}$-$v$ map by this standard deviation.

We repeated calculation of the weighted cross-correlation and the $K_\mathrm{p}$-$v$ map calculation for each of the 20 \texttt{SYSREM} iterations as well as for data with an injected signal. The best \texttt{SYSREM} iteration was then determined objectively, based on the highest signal to noise ratio recovery of the injected artificial signal analogous to \cite{18}. Details on the selection of the best \texttt{SYSREM} iteration can be found in Sect. \ref{sec:bestit}.% (Fig. \ref{fig:S3}).\\

\begin{table*}[ht]\renewcommand{\arraystretch}{1.5}
\centering
 \caption[]{Results from the cross-correlation analysis.}
 \begin{center}
\begin{tabular}{|l|c| cc | cc|}
 \hline \hline
  Molecule&  Best iteration & Blueshifted peak S/N & [$K_\mathrm{p}$, $v$] (km s$^{-1}$)&Redshifted peak S/N & [$K_\mathrm{p}$, $v$] (km s$^{-1}$)\\ \hline
H$_2$O & 7&8.7&[133, -8]&5.0&[136, +8] \\
CO & 5&9.6&[144, -8]&6.0&[143, +7] \\
CO +H$_2$O & 9&11.6&[140, -8]&6.6&[138, +8] \\
\hline
\end{tabular}
\tablefoot{The results listed are those obtained from the data that had undergone the \texttt{SYSREM} correction in stellar rest frame. The map was sampled in 1 km s$^{-1}$ steps which is reflected in the precision of the peak position coordinates.}
\end{center}
\label{tab:s3}
\end{table*}

\subsection{Results of the cross-correlation analysis}
\label{sec:cc_results}
For both H$_2$O and CO the CCF showed two spectroscopically resolved signals each. These signals were moving with the predicted planetary orbital velocity but shifted systematically by approximately +8~km~s$^{-1}$ and –8 km~s$^{-1}$, respectively, over the entire duration of the transit (Fig. \ref{fig:1} upper panel). This behaviour of the signals was also reproduced for the cross-correlation with a synthetic model containing both species. In each of the three cases, we found the amplitude of the redshifted signal to be lower than the blueshifted one. For CO$_2$ and CH$_4$ we found no significant signals (> 4$\sigma$) close to the expected $K_\mathrm{p}$.

\begin{figure}
\centering
\includegraphics[width=\hsize]{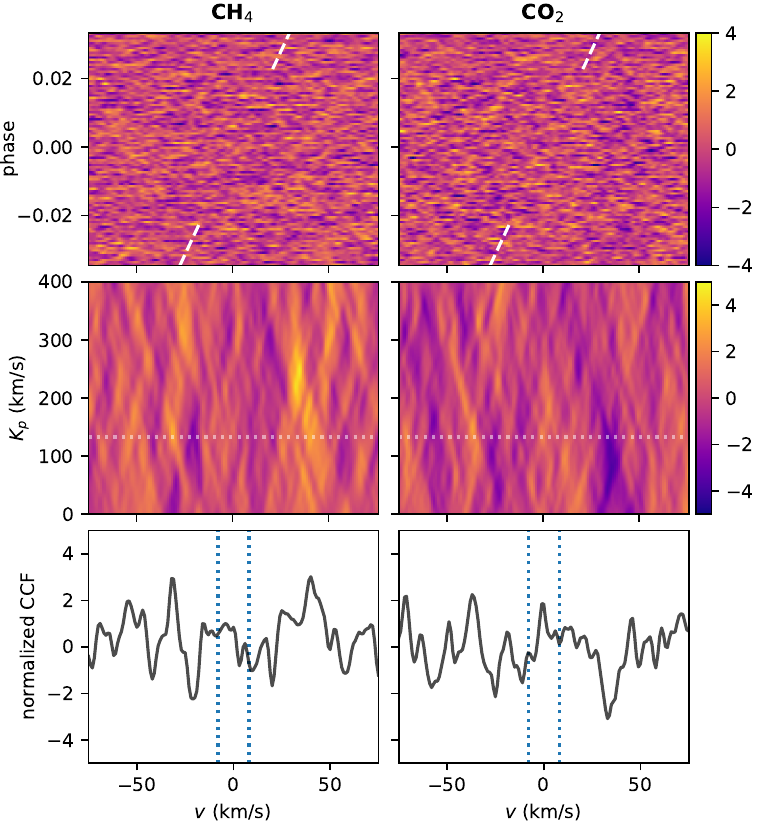}
  \caption{Results of the cross-correlation analysis for methane (CH$_4$, {left column}) and carbon dioxide (CO$_2$, {right column}). The results shown are based on data corrected by using \texttt{SYSREM} in telluric rest frame. {Top panels:} Cross-correlation function for each orbital phase. {Middle panels:} $K_\mathrm{p}$-$v$ maps. {Bottom panels:} Cut through the $K_\mathrm{p}$-$v$ map at the literature $K_\mathrm{p}$ value. }
     \label{fig:S5}
\end{figure}

The detections and non-detections as well as the position of the peak signals did not vary significantly between \texttt{SYSREM}  iterations. 
Furthermore, the double-peaked nature of the signals could be reproduced when confining our analysis to only spectra obtained in either nodding position A or B, as well as when only looking at a single wavelength segment (tested by using the last i.e. the reddest segment). Further, a cross-correlation of the synthetic models with themselves did not produce any other significant peaks beside the central one, with any aliasing effects being below 3.5\% of the central peak signal amplitude. We can, therefore, exclude that the signal shape was caused by co-adding misaligned data or by model self-correlation. The results are, moreover, robust against small variations in the model parameters. However, we noted that inclusion of a cloud or opacity deck (here at $\log_{10}(P_\mathrm{c}/\mathrm{bar})=-3$) increased the signal strength in the CCFs, which is why we choose to show the final results for the cross-correlation for this model. We show the cross-correlation results for the best \texttt{SYSREM} iteration in \mbox{Fig. \ref{fig:1}} for the detected species and in Fig. \ref{fig:S5} for the species without detected signals. The best number of \texttt{SYSREM} iterations and the corresponding S/N values and exact $K_\mathrm{p}$ and $v$ values of the peak signals are summarised in Table~\ref{tab:s3}.

We find that the two signals at $\approx\pm8$ km s$^{-1}$ peak at slightly different $K_\mathrm{p}$-values, which may be attributed to superimposed noise in the map or effects of \texttt{SYSREM} on the data, as the latter was not accounted for in the cross-correlation analysis. We find that a cut through the $K_\mathrm{p}$-$v$ map at either peak $K_\mathrm{p}$ value differs only marginally from a cut through the $K_\mathrm{p}$ value which we derive from literature parameters (i.e., $K_\mathrm{p}$ =133 km s$^{-1}$). Consequently, in the bottom panels of Figs. \ref{fig:1}, \ref{fig:S4}, and \ref{fig:S5}, we show a cut through the $K_\mathrm{p}$-$v$ map at $K_\mathrm{p}$ =133 km s$^{-1}$.

\subsection{Discussion of the cross-correlation results}\label{sec:cc_discussion}
We confirmed that the signals of H$_2$O and CO align at $K_\mathrm{p}$ velocities compatible with the prediction from the literature values, supporting the conclusion that they are of planetary origin and that both species are present in the planet atmosphere. The absence of signals of CH$_4$ and CO$_2$, however, does not necessarily imply that these species are not contained within the atmosphere at all. When computing a model spectrum containing the two detected species and the two non-detected species, we found that this model is indistinguishable from a model only containing the two detected species. The explanation for this is that the absorption features of water are forming a quasi-continuum that hides any weaker absorption features, for example those of CH$_4$ and CO$_2$ in the case of solar metallicity and C/O ratio.

Furthermore, we found indications of the presence of an additional opacity source, as the obtained S/N  of the  signals increased when  an additional absorber was introduced in the form of a grey opacity or cloud deck. The presence and exact pressure level of such a potential absorption layer as well as the abundances of atmospheric species, however, cannot be accurately determined by testing a limited number of models in cross-correlation. Instead, a Bayesian retrieval framework can be employed. In this framework we   have to address the fact that our planet signal is spread over different regions in velocity space. The double-peaked nature of our signals suggests that the atmosphere of this planet cannot be accurately characterised by a 1D model. To adapt to a physically more accurate description of the planet signal we must understand what physical circumstances could cause the  signal of the planet's transmission spectrum to be split into two distinct peaks. In the following section we discuss the double-peaked signal and its probable source.

\subsection{Discussion of the double-peaked velocity signal}\label{sec:doublepeak}
\begin{figure*}
\centering
\includegraphics[width=\hsize]{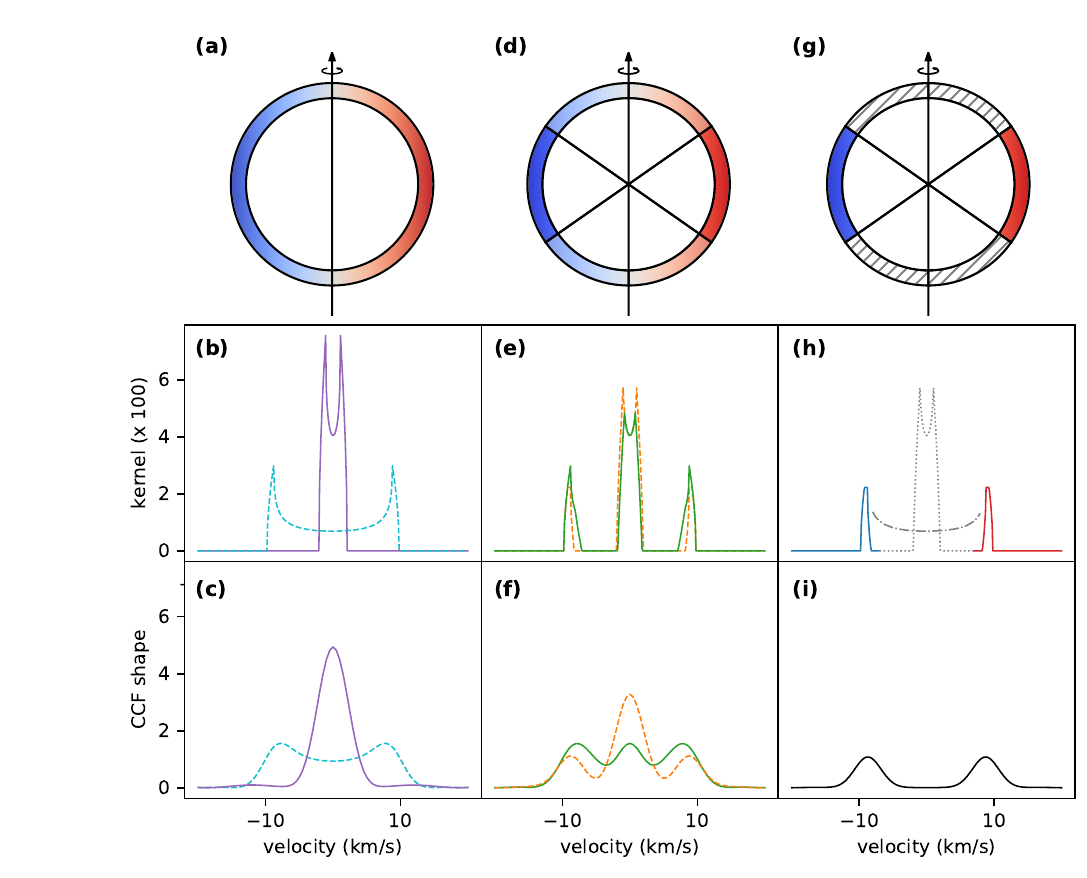}
  \caption{Different scenarios of super-rotating atmospheric material.  {Panels a, b, and c} show the scenario of a homogeneously rotating planet. The purple solid line shows the velocity kernel of the material at rotation consistent with tidally locked rotation ({panel b}) and the resulting shape of the cross-correlation function ({panel c}). The turquoise dashed line shows the effect of the planet rotating approximately six times faster. {Panels d, e, and f} show two cases of a super-rotating equatorial jet, with the poles rotating at tidally locked velocities. The difference between the two cases is the opening angle of the equatorial region (35$^\circ$ orange dashed, and 65$^\circ$ solid green). {Panels g, h, and i} explore the case of the velocity kernel of a super-rotating equatorial jet (blue and red solid lines in {panel h}) with the polar region represented by the velocity kernel of super-rotating velocity (grey dash-dotted line) or tidally locked velocity (grey dotted line), but not contributing to the transmission spectrum due to muted atmospheric features, leading in both cases to only two peaks in the CCF ({panel i}). }
     \label{fig:2}
\end{figure*}
The exoplanet WASP-127\,b is a hot Jupiter orbiting a G-type star at 0.05 AU distance on a misaligned orbit with a 4.2-day period \citep{22, 23}. At this distance to its host star, the planet is presumed to be tidally locked, meaning that it completes one rotation around itself during each orbit. Given the radius of the planet of 1.3 Jupiter radii \citep{24}, this would lead to a maximum velocity at the equatorial regions of $v_\mathrm{max}= 1.6$ km s$^{-1}$.

Figure \ref{fig:2} shows different example scenarios for atmospheric dynamics and their influence on the velocity kernel of the atmospheric material. The top row of Fig. \ref{fig:2} (panels a, d, g)  shows sketches of the described scenarios. The middle row (panels b, e, h) shows the velocity kernels of the material at infinite resolution. These kernels were calculated as described in Sect. \ref{sec:velocityprofilecalculation}. The bottom row (panels c, f, and i) shows how the CCF of an atmosphere signal that was broadened with the velocity kernel from the middle row would appear in observations at a resolution of $\mathcal{R}= 140\,000$. For this we used a synthetic atmospheric model, which we first  convolved with the velocity kernel depicted in the middle row, and then convolved with a Gaussian to apply the effects of instrument resolution as  described in Sect. \ref{sec:resolutionnorm}. Then we cross-correlated this broadened model with its unbroadened counterpart to create the CCFs shown in the bottom panels. In the scenario for the muted polar regions (g, h, i), a featureless model was used for the convolution with the part of the velocity kernel representing the polar regions, in all other cases we used the model employed in the CCF calculations of Sect \ref{sec:1dmodels}. In the following we address the different scenarios in detail. Figure \ref{fig:2} b shows the velocity distribution of the material in the part of the atmosphere probed by transmission spectroscopy during transit for the case of tidally locked rotation. While this distribution exhibits two distinct peaks they are very close in velocity space. Due to the limited resolution of the instrument, this scenario leads to a broadened cross-correlation function instead of a double-peak profile (Fig. \ref{fig:2} c, solid line). The two peaks of the velocity kernel   only become spectrally resolved in the CCF at higher rotational velocities (Fig.~\ref{fig:2}~c, dashed line;  also see e.g. \citeauthor{25}, \citeyear{25}), at which point the CCF will show two peaks representing the material at the two equatorial terminator regions and a plateau between them that is caused by the decrease in  the velocities when moving towards higher latitudes (i.e. closer to the rotation axis).

An alternative scenario does not require the entire planet to rotate at this elevated speed but constrains these fast velocity regions to the equator, where they form an equatorial super-rotating jet (Fig. \ref{fig:2} d).  Such jets are predicted by 3D models to occur in hot Jupiter planets  \citep{26, 27} and have been suggested from line broadening in high-resolution planetary dayside spectra \citep{28, 29}, ingress and egress asymmetries in transmission spectroscopy \citep{7, 30}, and hot spot shifts derived from low-resolution eclipse mapping \citep{31}.
In the case of the fast-rotating material being constrained to a region at the equator, the remaining signals at the poles, rotating at reduced velocities in agreement with tidally locked rotation, would form a third peak at around 0 km s$^{-1}$ (Fig.~\ref{fig:2}~e,~f).

As we do not detect this peak, nor a plateau between our two signals at $\pm8$ km s$^{-1}$, we hypothesise that the signal of the planet atmosphere in the polar region of WASP-127\,b may be significantly muted due to lower temperatures, a higher cloud deck, or a combination of these factors. Similarly, these parameters being different between the evening terminator (i.e. where the material transitions from the day to the nightside and is blueshifted) and the morning terminator (where the material moves from the nightside to the dayside and is redshifted) could explain the difference between the amplitudes found for these two signals in the CCF. We evaluate whether such a scenario could produce two isolated peaks by assigning a flat model spectrum to the polar regions of the planet. Figure \ref{fig:2} i shows that for this scenario the resulting CCF signal only exhibits two peaks without a plateau or third peak at low velocities, which is compatible with our findings for WASP-127\,b.

\section{Two-dimensional retrieval of the atmosphere}
\label{sec:retrievalsetup}
We set up a retrieval framework using a log-likelihood optimisation \citep[as described in][]{2020A&A...640L...5Y,32,29,2024A&A...688A.206C}, using the Markov chain Monte Carlo (MCMC) method, as implemented by \cite{33}. 
We adapted this retrieval framework by switching from a 1D atmospheric model to a simplified 2D model. We assumed that the probed part of the atmosphere in transmission can be described as a ring around the planet. Instead of using one 1D model to represent the entire ring, we divided it into three regions, with each region represented by its own 1D model (calculated with \texttt{petitRADTRANS}). Similar 2D retrieval approaches have been employed by previous studies \citep{6, 2023AJ....165..242G, 36,2024A&A...687A..49M}, where the atmospheric ring was split into two regions representing the signals of the leading and trailing planetary limb observable at a given time. 
In our case the three regions represent the two equatorial regions at the leading and trailing terminators and the polar regions (combined as one region) as indicated in Fig. \ref{fig:2} g. The model of each of the three regions was convolved with the respective velocity distribution (calculated as described in Sect. \ref{sec:velocityprofilecalculation} and indicated in Fig. \ref{fig:2} h) and the weight of its corresponding region, before all models were combined into a global model. The global model was then shifted into the rest frame of the data for each exposure and the effects of \texttt{SYSREM} on the data were reproduced in the model following the pre-processing approach described by \cite{46}.

For the comparison of the model with the data, we used a Gaussian log likelihood function following \cite{17}, \cite{2020A&A...640L...5Y,32}, \cite{29} and \cite{2024A&A...688A.206C}: 
\begin{equation}
\label{eq:likelyhood}
    \ln{L} = -\frac{1}{2}\sum_{i,j} \left[ \frac{\left(R_{ij} - M_{ij}\right)^2}{\left(\beta S_{ij}\right)^2} + \ln{2 \pi \left(\beta S_{ij}\right)^2} \right] \mathrm{.}
\end{equation}
Here $R$ is the 2D matrix of the residual spectra, $S$ is the matrix containing the corresponding errors, $M$ is the matrix of the pre-processed model, and $i$ and $j$ indicate the individual elements of the 2D data. We also allow for an error scaling factor $\beta$. In the following we discuss the various free parameters of the retrieval, regulating the velocity distributions, atmospheric conditions in the three regions, and transformation into data rest frame.
\subsection{Free parameters}
\subsubsection{Parameters regulating the velocity profile}
As the latitudinal angle $\varphi$ at which an equatorial region transitions into the polar region was unknown, it was set as a free parameter in the retrieval. In  Fig. \ref{fig:2} g, as one
possible example, it is set to $\varphi = 35^\circ$  (see Fig. \ref{fig:3} a for an illustration with the retrieved value $\varphi = 65^\circ$). Further free parameters regulating the velocity profile were the maximum equatorial velocity $v_\mathrm{eq}$ for the super-rotating material in the equatorial regions and a parameter regulating the velocity at the polar region $v_\mathrm{p}$. Here $v_\mathrm{p}$ does not correspond to the maximum velocity measured for the polar regions, but rather the maximum velocity that would be measured at the equator if the entire planet was rotating at the same velocity as the polar regions. The maximum velocity can be translated into an angular velocity as $\omega = v/R_\mathrm{p}$ (s$^{-1}$). While we initially considered fixing the maximum velocity for the polar regions, $v_\mathrm{p}$, to that corresponding to tidally locked rotation, we ultimately did not want to constrain the validity of the results by this assumption, and we decided to allow the value to vary freely within the limits of a uniform prior, covering values between the tidally locked case (1.6 km s$^{-1}$) and the faster velocity obtained for the super-rotating equatorial material $v_\mathrm{eq}$. This allowed us to obtain more accurate uncertainties for the atmospheric parameters of this region. The formal calculation of the velocity kernel is described in Sect. \ref{sec:velocityprofilecalculation}.

\subsubsection{Parameters regulating atmospheric properties}\label{sec:retrievalchem} 
The atmospheric chemistry was parametrised via one joined value for the C/O ratio and one for the metallicity, both of which   were shared between all three modelled regions. Aside from the different velocity kernels, the atmospheres representing the three regions were parametrised as distinct from each other by using individual cloud top pressures $P_\mathrm{c}$ and individual  \mbox{$T$-$p$ profiles}. 

We find that Guillot \mbox{$T$-$p$ profiles}, with values for $\kappa_\mathrm{IR}$, $\gamma$, and $T_\mathrm{int}$ fixed to the values obtained by \cite{21} and varying $T_\mathrm{eq}$, approach isothermal \mbox{$T$-$p$ profiles} at pressures below  approximately $\log_{10}(P_c)=-2$. We further find that the shape of the temperature profiles below this pressure level have very little impact on the resulting models. This impact gets further minimised in the presence of cloud or opacity decks. This effect can be explained when we   look at the contribution function for a few sample models at different $T_\mathrm{eq}$.   
Here we see that the majority
 of transmission signal originates from one specific pressure level with only minor contributions from pressures lower than this level and no contributions from pressures higher than this level even in the absence of clouds. This suggests that our data is very sensitive to a particular pressure level, but not sensitive to the potentially more complex shape of the $T$-$p$ profile at higher pressures. This means that the data does not contain the necessary information to constrain the parameters of $\kappa_\mathrm{IR}$, $\gamma$, and $T_\mathrm{int}$. Instead, we are mostly sensitive to the specific temperature value of the $T$-$p$ profile in the pressure region from which the bulk of the contribution to the transmission spectrum originates. An isothermal profile adopting the temperature of that pressure level thus appears to be an adequate simplification for the $T$-$p$ profiles in our retrieval.

To further assess whether the choice of an isothermal $T$-$p$ profile is an appropriate assumption, we tested our retrieval with three different parametrisation approaches: I) with isothermal \mbox{$T$-$p$ profiles}; II) with Guillot \mbox{$T$-$p$ profiles,} where $T_\mathrm{eq}$ is a free parameter and $\kappa_\mathrm{IR}$, $\gamma$, and $T_\mathrm{int}$ are fixed to the values given by \cite{21} (an approach  analogous to that used by \citealt{36}); and     III) Guillot \mbox{$T$-$p$ profiles} with all four parameters set as free parameters. We obtain identical results for all parameters using approaches I and II. The results of approach III confirm that our data are not sensitive to $\kappa_\mathrm{IR}$, $\gamma$, and $T_\mathrm{int}$ as these values are poorly constrained by the retrieval, which does not appear well converged even after 10\,000 steps. Nonetheless, the results obtained for all remaining atmospheric and dynamic parameters are still compatible with the results from I and II. As the assumptions of $\kappa_\mathrm{IR}$, $\gamma$, and $T_\mathrm{int}$ in II appear to have no impact on the results, we opted to proceed with the isothermal \mbox{$T$-$p$ profile} parametrisation. In this case, the retrieved value can be interpreted as the  temperature that the likely more complex real \mbox{$T$-$p$ profile} adopts in the pressure regime that contributes most significantly to the spectral transmission signal.\\
While the simplified illustration in Fig. \ref{fig:2} g does not explicitly depict any change in planet radius between the three different regions, this effect is incorporated in the modelling by the possibility that the three regions are characterised by different temperatures. For example, by increasing the temperature for the 1D model representing the evening terminator with respect to that of the morning terminator, the scale height of the former model would increase, and the 100 atmospheric pressure layers computed by \texttt{petitRADTRANS} would be spread over a wider altitude range. This would then result in the typical asymmetric planet radius often depicted in other works discussing terminator differences \citep[e.g.][]{44, 45}.

\subsubsection{Other parameters}
The remaining free parameters were the velocity semi-amplitude of the planet orbit $K_\mathrm{p}$ as well as a velocity shift with respect to the literature systemic velocity $v_0$. The two parameters were required to compute the shift of the model from planet rest frame into the rest frame of the data.  Finally, we  allowed for a noise scaling factor $\beta$ to address possible under- or overestimation of the data errors in the calculation of the likelihood (see Eq. \ref{eq:likelyhood}).

\subsection{Set-up of the MCMC}
All the parameters were constrained by uniform priors. In each case but one we used a very large interval that did not constrain the posterior distribution. The exception was the case of $v_\mathrm{p}$, which, as mentioned before, was constrained between the tidally locked case and the equatorial maximum velocity. This was done because the data did not contain any information on this parameter, and it was only allowed to vary within the reasonable possible parameters to yield more conservative uncertainties for other possibly correlated parameters, for example the temperature and cloud top pressure of the polar region. The MCMC was set up with 32 walkers and was run for 11\,000 steps. We determined the burn-in as the point at which adding more steps to the chain did not alter the results. We identified this point by two means. First, we visually inspected the chains to determine the moment all chains had converged, and second, as a more objective approach, we calculated the mean and 1$\sigma$ uncertainty intervals of all values and determined the burn-in step as the step from which on these values did not significantly change any more, regardless of how many following steps we included in the calculation. Both methods led us to the conclusion that the chains were only fully converged after 5000 steps. We thus disregarded the first 5000 steps of the chain and used   steps 5001 to 11\,000 to calculate the results.

\subsection{Calculation of the velocity profiles based on super rotation and equatorial jet}\label{sec:velocityprofilecalculation}
We propose that the probed region of the atmosphere, the terminator, can be described by a ring around the planet (as illustrated in Fig. \ref{fig:2}~a,~b). For the computation of the velocity kernel, we chose a numerical approach. We  divided the ring into 36\,000 cells of equal weight. The velocity of the material represented by each cell depends on the cell’s distance from the planet’s rotation axis. Each cell can be identified by its unique angle $\vartheta$, where $\vartheta$ spans the range from 0$^\circ$ to 360$^\circ$ in 36\,000 steps of 0.01$^\circ$. The velocity of the cell corresponding to $\vartheta$ is then calculated as $v_\mathrm{kernel}(\vartheta)=v_\mathrm{max} \sin(\vartheta)$. In this equation, $v_\mathrm{max}$ is the maximum velocity the material would have at the point farthest away from the rotation axis (i.e. at the equator). For a planet with material rotating homogeneously at all latitudes, this corresponds to the velocity of the material at the equator $v_\mathrm{eq}$. However, for a planet where the equatorial regions rotate faster than the poles, the pole velocity kernel may be described with a different $v_\mathrm{max}$ that is smaller than the actual velocity at the equator, for example with the one corresponding to tidally locked rotation ($v_\mathrm{tidally~locked}$= 1.6 km s$^{-1}$ in the case of WASP-127\,b).\\ 
We then  computed the kernel numerically by calculating the velocity for each angle and building a histogram of the sampled velocities. We note that this approach only considers one atmospheric layer (i.e. one cell per angle element) that spans the entire height of the atmosphere, for which the  mean velocity of the cell is adopted (representative for the centre  of the cell). We experimented with making this model more complex by adding several layers, resulting in multiple rings around the atmosphere, but found that the added complexity only resulted in an additional broadening of the resulting velocity profile, which becomes negligible when also applying instrument broadening to the data. The introduction of several layers would, however, require us to make assumptions about the planet radius and the width of the atmospheric annulus, whereas the one-layer model is independent of such assumptions. 
Furthermore, applying altitude dependent velocity shifts in a multi-layer atmosphere correctly would require different parts of the 1D model to be shifted by different velocities  (e.g. shifting the line cores forming high in the atmosphere, and thus  farther away from the rotation axis by larger velocities than the line wings that form  in deeper layers, and thus closer to the rotation axis). Therefore, we refrain from using multiple atmospheric layers and proceed here with the simplified one-layer assumption. We find that the velocity kernels calculated with this approach yield equivalent results to those calculated with  the analytical approach of \cite{6,2023AJ....165..242G}, and those of \cite{36} and \cite{2024A&A...687A..49M} if for the latter the width $h$ of the atmospheric ring is assumed to be small compared to the planet radius ($h<< R_\mathrm{p}$). \cite{36} and \cite{2024A&A...687A..49M} use the assumption of a fixed annulus width of $h=5 H$ in their work. At our instrument resolution of $\mathcal{R}=140\,000$ we find that our kernels are compatible to theirs for annulus widths up to $h=0.25 R_\mathrm{p}=10 H$ (with the atmospheric scale height of WASP-127\,b being $H\approx2\,000$ km).

Since we do not consider the height of the atmosphere in the calculation of the velocity kernel in general, we also do not consider possible differences in the height of individual cells. However, as stated in Sect. \ref{sec:retrievalchem}, the freedom of different atmospheric (scale) heights is still given to the model within the retrieval, where different clusters of cells can be assigned different model atmospheres (with different $T$-$p$ profiles and cloud top pressures, effectively leading to more extended or less extended atmosphere heights). Lastly, we allow different parts of the ring to rotate around the common rotation axis at different velocities by setting individual values for $v_\mathrm{max}$ for these parts of the ring. Specifically, we separate a region around the planet equator from the higher latitudes.

The point at which one region ends and another begins is parametrised by the angle $\varphi$, where the regions of the ring described by the conditions ($90^\circ- \varphi) \leq \vartheta  \leq (90^\circ+ \varphi)$ and $(90^\circ+180^\circ -\varphi) \leq \vartheta  \leq (90^\circ+180^\circ+ \varphi)$ are considered the equatorial regions. The first term describes the equator at the redshifted terminator rotating away from the observer and the second describes the equator of the blueshifted terminator rotating towards the observer. Consequently, the super-rotating equatorial regions have an opening angle of $2\varphi$, centred around the exact equatorial regions at $\vartheta  = 90^\circ$ and $\vartheta  = 270^\circ$.  
\begin{figure}
\centering
\includegraphics[width=\hsize]{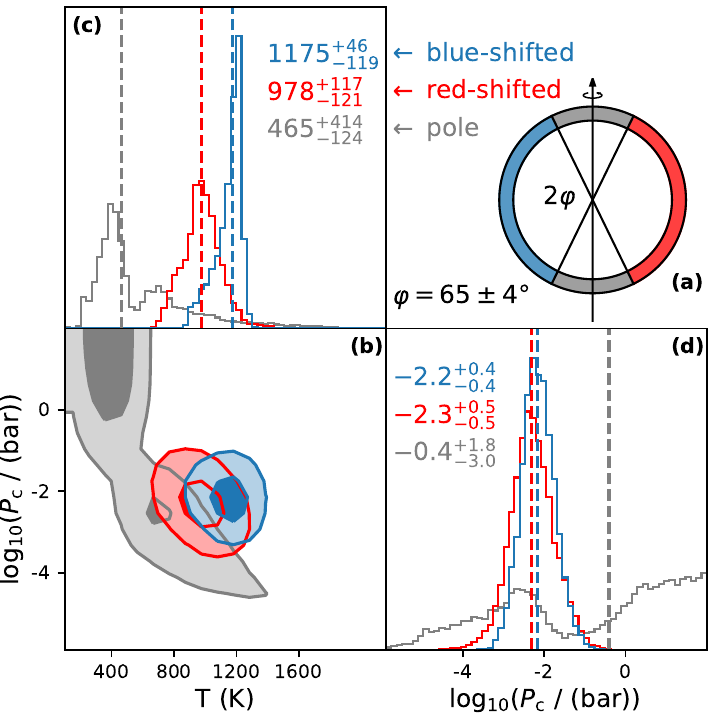}
  \caption{Atmospheric properties of the three different modelled regions of the planet. {Panel a:} Sketch of the three different regions retrieved for the planet atmosphere.  {Panel b:} Probability distributions of the temperature and cloud top pressure retrieved for the three regions. The isolines indicate the 1$\sigma$ and 2$\sigma$ levels of a 2D distribution, i.e. encompassing 39.3\% and 86.0\% of all posterior samples, respectively. The blueshifted terminator region is indicated in blue, the redshifted terminator region is indicated in red, and the polar region is indicated in grey. {Panel c:}  Histograms of the temperature distributions. {Panel  d:} Histograms of the cloud top pressure distributions. The mean values of the distributions are indicated with a dashed line. The respective values are quoted in the plot, together with the 1$\sigma$ uncertainties of a 1D distribution, i.e. determined as the values that encompass 68.0\% of the distribution.} 
     \label{fig:3}
\end{figure}
\begin{figure}
\centering
\includegraphics[width=\hsize]{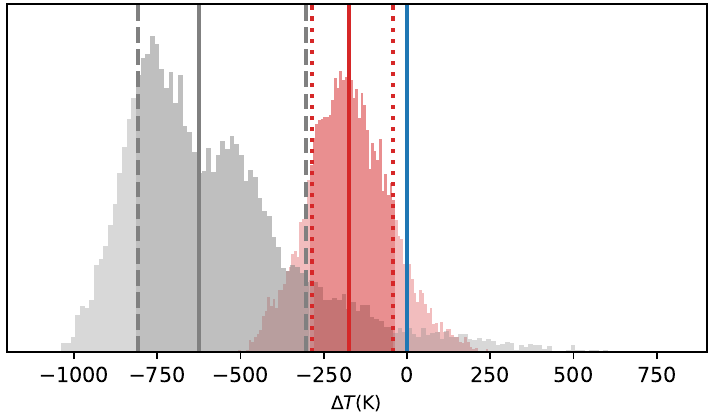}%plots_AAdelta_T_v3.pdf}
  \caption{Histogram of the differential temperature distributions. In red we show the distribution of the temperature difference of the redshifted region to the blueshifted region $\Delta T_\textrm{red} = T_\textrm{r} -  T_\textrm{b}$. In grey we show the temperature difference of the polar regions with respect to the blueshifted region $\Delta T_\textrm{pole} = T_\textrm{p} - T_\textrm{b}$. The red and grey solid lines   indicate the median of the respective distributions. The blue solid line   indicates the zero point as the position of the relative temperature of the blueshifted region. The regions between the 16th and 84th percentiles, indicative of the 1$\sigma$ uncertainties of each of the two distributions, are indicated in deeper colours and additionally by dashed grey and dotted red lines for the pole and redshifted distributions, respectively. The corresponding values are $\Delta T_\textrm{red} = -175^{+133}_{-116}$ K and $\Delta T_\mathrm{pole} =-624^{+328}_{-190}$ K. While both distributions overlap, their peaks are distinct from each other and distinct from zero by approximately $1\sigma$ and $1.9\sigma$, respectively.}
     \label{fig:4}
\end{figure}

\subsection{Injection-recovery test}
In order to verify that our retrieval is able to correctly constrain the atmospheric parameters of different regions on the planet and our retrieved values are not strongly biased by the data pre-processing, we performed an injection recovery test. In this test we injected a synthetic model into the data shifted by -60 km s$^{-1}$ with respect to the real signal. We injected the model at the very beginning of our analysis, before normalisation. The injected model contains H$_2$O and CO as well as an opacity deck. 
We find that the retrieved values correspond to those of the injected model within the 2$\sigma$ uncertainties for all parameters and within 1$\sigma$ for 9 out of the 14 free parameters. 
Among the values retrieved correctly within the 1$\sigma$ uncertainties are the  C/O ratio and metallicity as well as the temperatures of the morning terminator and the poles. 
While the  absolute temperature of the evening terminator is only retrieved within 2$\sigma$ of the injected value, it is still retrieved accurately within 100 K. Moreover, the temperature differences between each of the three regions are also correctly retrieved within 1$\sigma$.

We note that all temperatures are systematically slightly overestimated, while the corresponding cloud deck pressures are retrieved as slightly too low, with the posteriors of the two parameters showing  correlation. The MCMC thus appears to converge at a slightly different than the injected combination of the two parameters, which is also able to explain the signal. This highlights how high-resolution studies can benefit from low-resolution complementary information, which can help break this degeneracy through measurements of the true continuum level.

\subsection{Results of the retrieval and discussion}\label{sec:retrieval_results_discussion}
The resulting posterior distributions of the free parameters are shown in Fig. \ref{fig:S6}. Also quoted are the mean values and the 16\% and 84\% quantiles as 1$\sigma$ uncertainty intervals. For some values, we find that the mean does not correspond to the most sampled values, for example the distributions are very asymmetric. We  also calculate the mode of the distributions (i.e. the most frequently sampled value). To obtain the mode, we round all values of a specific parameter to a lower precision (i.e. 10\,K for the temperatures) and then calculate the most frequently sampled values from these rounded values. We quote the prior limits of the free parameters, the mean, the 1$\sigma$ uncertainties, and the mode of their posterior distributions, as well as the precision to which the values were rounded prior to the mode calculation in Table \ref{tab:rerivalresults}.\\
\begin{table*}[ht]\renewcommand{\arraystretch}{1.5}
\centering
 \caption[]{Results of the atmospheric retrieval.}\label{tab:rerivalresults}
\begin{tabular}{llllll}
 \hline \hline
  Parameter &
  Prior range &
  Median value &
  Mode&
  Mode precision&
  Unit
 \\ \hline
$T_\mathrm{b}$   & [100, 3000] & 1174.97$^{+49.07}_{-119.25}$ & 1210&10& K  \\
$\log_{10}(P_\mathrm{c, b})$ & [$-5.9$, 2] & $-2.16^{+0.42}_{-0.38}$  &$-2.3$&0.1&  $\log_{10}$(bar)  \\
$T_\mathrm{r}$   & [100, 3000] & 977.67$^{+117.38}_{-120.94}$ & 950&10&  K  \\
$\log_{10}(P_\mathrm{c, r})$ & [$-5.9$, 2] & $-2.31^{+0.53}_{-0.49}$  &  $-2.4$&0.1&$\log_{10}$(bar)  \\
$T_\mathrm{p}$   & [100, 3000] & 465.01$^{+413.54}_{-123.81}$ & 410&10& K  \\
$\log_{10}(P_\mathrm{c, p})$ & [$-5.9$, 2] & $-0.41^{+1.76}_{-2.96}$  & 1.9&0.1& $\log_{10}$(bar)  \\
C/O & [0,1.5] & $-0.56^{+0.05}_{-0.07}$  & 0.56 & 0.01 &  ... \\
$[$Fe/H$]$& [$-5$, 5] & $0.01^{+0.36}_{-0.39}$  &0.07&0.01& ...  \\
$v_\mathrm{eq}$   & [1, 15] & $9.26^{+0.17}_{-0.21}$  & 9.4&0.1& km\,s$^{-1}$  \\
$v_\mathrm{p}$   & [1.6, $v_\mathrm{eq}$] & $5.12^{+2.69}_{-1.39}$ & 4&1&  km\,s$^{-1}$  \\
$\varphi$& [1, 89] & $64.61^{+3.75}_{-3.98}$  & 65&1& $^\circ$  \\
$K_\mathrm{p}$   & [70, 230] & $133^{+2.02}_{-1.88}$  &  133.6&0.1&km\,s$^{-1}$  \\
$v_0$   & [-6, 6] & $-0.40^{+0.14}_{-0.14}$ &  $-0.4$&$0.1$& km\,s$^{-1}$  \\
$\beta$   & [0.0001,1000] & $0.60894^{+0.0004}_{-0.0004}$  & 0.609&0.001& ...\\
\hline
\end{tabular}
\tablefoot{For the temperatures $T$ and cloud top pressures $\log_{10}(P_\mathrm{c})$ the index b stands for blueshifted signal, r for redshifted signal, and p for polar regions. The uncertainties quoted with the median value correspond to the 16th and 84th percentiles. We also list the mode of each distribution (with the distribution values rounded to the precision quoted).}
\end{table*}
\indent The retrieval confirms that the data can be explained by super-rotation of the atmospheric material. We retrieve a maximum velocity at the equator of $v_\mathrm{eq} = 9.3 \pm 0.2$ \mbox{km s$^{-1}$}. This is approximately six times faster than velocities resulting from tidally locked rotation. It is well below the escape velocity \mbox{($v_\mathrm{escape}$ = 21.3 km s$^{-1}$),} but above the speed of sound.  The latter is positively correlated with temperature and can be estimated for the hot conditions of a heated exoplanet atmosphere using Eq. 1 from \cite{2022AJ....163..155P}, who observed supersonic day-to-nightside wind speeds for the ultra-hot Jupiter KELT-9\,b. Using their formula we obtain sound speeds of the order of 3 \mbox{km s$^{-1}$} for WASP-127\,b.
Under the assumption of a tidally locked rotating planet, the retrieved equatorial velocity corresponds to an excess wind speed of 7.7 km s$^{-1}$, and thus to the material being transported from one terminator to the other in 10.4 hours and revolving around the entire equator in 0.87 days. For comparison, the maximum equatorial wind speeds for Saturn and Jupiter are \mbox{0.43 km s$^{-1}$} and \mbox{0.14 km s$^{-1}$}, respectively \citep{34}.
The latitudinal angle at which the equatorial zone with super-rotating material transitions into the polar zone is retrieved at $65\pm4^{\circ}$. This value aligns with general circulation models (GCMs), which depict that the jets  occupy similarly large areas of the planet \citep{35}.

The retrieval shows that the velocities at the polar region cannot be constrained from the data, likely due to lack of features detected. This manifests as the posterior distribution homogeneously covering the entire parameter space permitted by the prior. The lack of signal is translated into constraints on the temperature and cloud top pressure compatible with this result. Figure \ref{fig:3} shows the posterior distributions for the temperature and cloud top pressure for the three modelled regions: the morning and evening terminators, and the polar regions. We find that the retrieved values for the polar region occupy a distinct region in this 2D parameter space separated from the distribution of the  terminators by more than 1$\sigma$. We find that both terminators have compatible cloud top pressures $(\log_{10}(P_\mathrm{c,blue}/\mathrm{bar}) =  -2.2$ and $\log_{10}(P_\mathrm{c,red}/\mathrm{bar}) = -2.3)$, but peak at different temperatures. The distributions of the temperatures of the redshifted morning and the blueshifted evening terminator overlap within their 1$\sigma$ uncertainties. However, these uncertainties reflect the degeneracy of the absolute temperatures with other parameters of the retrieval, which are shared between the three regions and affect all temperature values systematically. Instead, the temperature differences between the individual regions in each sampled MCMC step allow  us to eliminate this effect and gives us a more robust measure of how distinct these regions are from each other.
\begin{figure}
\centering
\includegraphics[width=\hsize]{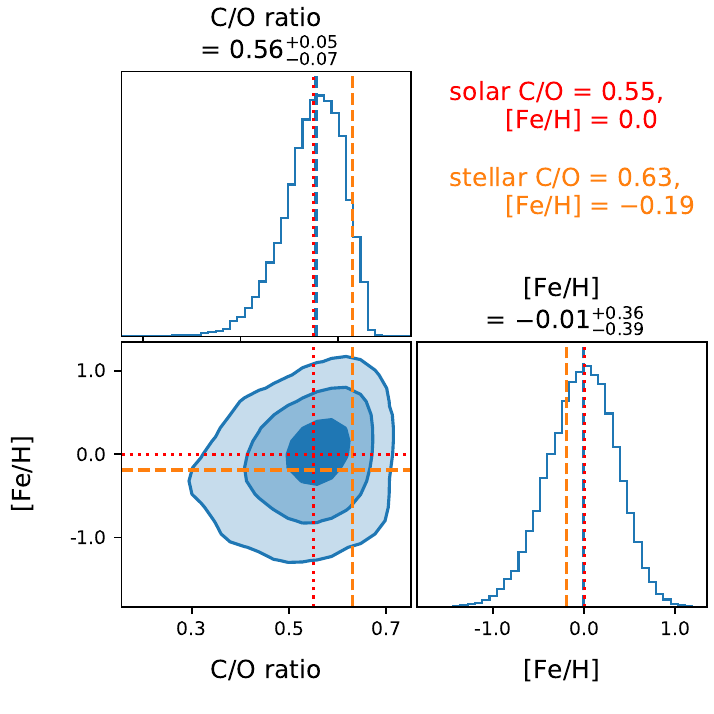}
    \caption{Posterior distributions for the C/O ratio and metallicity of the planet atmosphere. {Bottom left:} Posterior distributions from the retrievals for the C/O ratio and metallicity [Fe/H] of the planet (in blue). The isolines encompass 1$\sigma$, 2$\sigma$, and 3$\sigma$. Solar values are indicated by a red dotted line and the values obtained for the host star (\cite{36} are depicted by an orange dashed line. The median values of the planet value distributions are indicated with a dashed blue line in the histograms for the C/O ratio ({top left}) and the metallicity [Fe/H] ({bottom right}) and quoted above the panels with the respective histograms together with the 1$\sigma$ uncertainties. The dashed blue line for the median values falls nearly on top of the red dotted lines indicating the solar values. }
     \label{fig:S7}
\end{figure}
The resulting distributions of the temperature differences between the pole and blueshifted evening terminator ($\Delta T_\mathrm{pole} =-624^{+328}_{-190}$ K) and the redshifted morning terminator to the blueshifted evening terminator ($\Delta T_\mathrm{red} = -175^{+133}_{-116}$ K) are shown in Fig. \ref{fig:4}, and show tentative evidence for physically relevant differences in temperature of these three regions.
The higher temperature of the evening terminator compared to the morning terminator as well as the cooler temperatures obtained for the poles agree in general with predictions by  the GCMs of planets with super-rotating equatorial jets, where the hotter dayside material enters the nightside on the evening terminator and then cools down before exiting it on the morning terminator, while the higher latitudes do not mix significantly with the heated equatorial material and exhibit overall lower temperatures \citep{35}. We caution that our results are based on a retrieval in which the atmosphere in these regions is only permitted to vary in isothermal $T$-$p$ profiles and opacity deck pressures, while other parameters such as those regulating the chemical composition were kept fixed. It is possible that the significance of the temperature differences could be reduced if the model were given the freedom to explain the terminator differences through a combination of temperature and further independent parameters.\\\\
\begin{figure}
\centering
\includegraphics[width=\hsize]{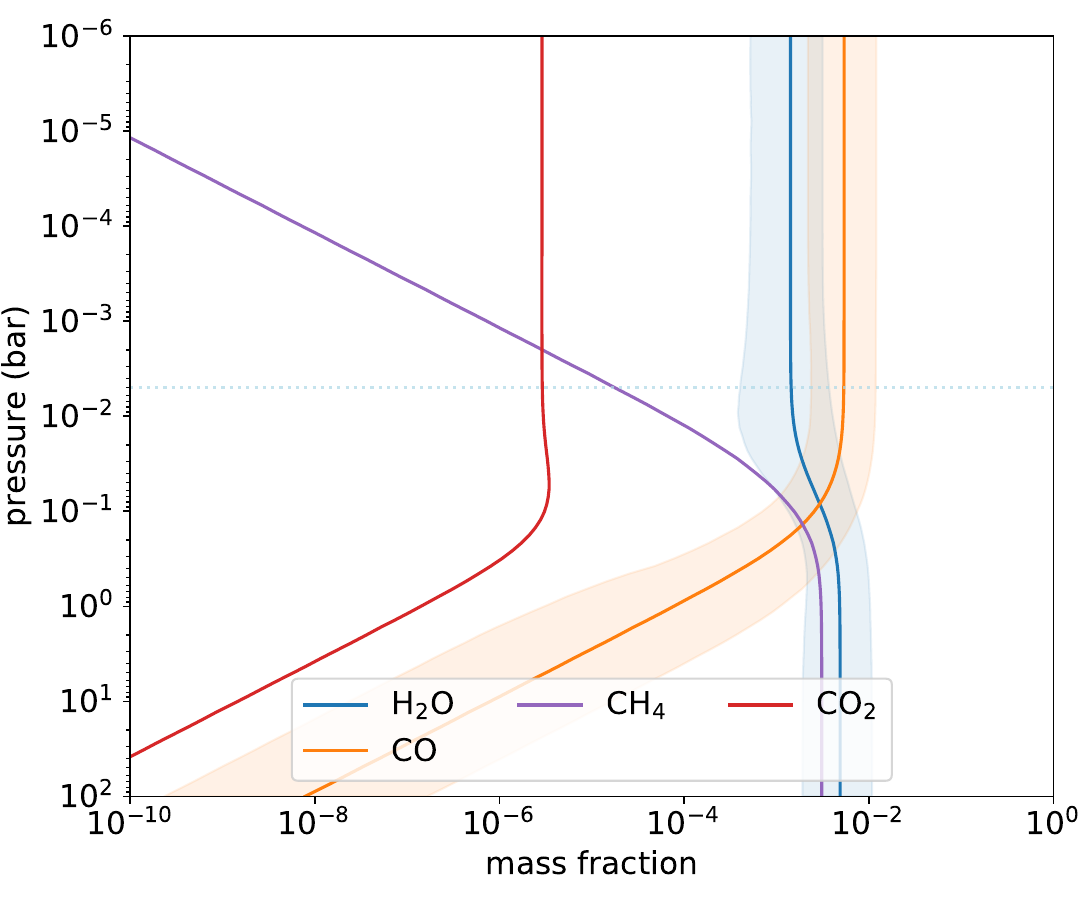}
  \caption{Mass fraction abundances for chemical equilibrium assumptions for the retrieved morning terminator model. For the calculation we used the retrieved C/O ratio, metallicity, isothermal temperature, and cloud top pressure of the redshifted (morning terminator) signal. We only show the uncertainties of the abundances of H$_2$O and CO as they were the only species detected in the data. We indicate the height of the cloud top pressure as a horizontal dotted light blue line. Our data are only sensitive to signals and the corresponding abundances above the cloud top pressure (i.e. above that line).}
     \label{fig:S8}
\end{figure}
\begin{figure}
\centering
\includegraphics[width=0.88\hsize]{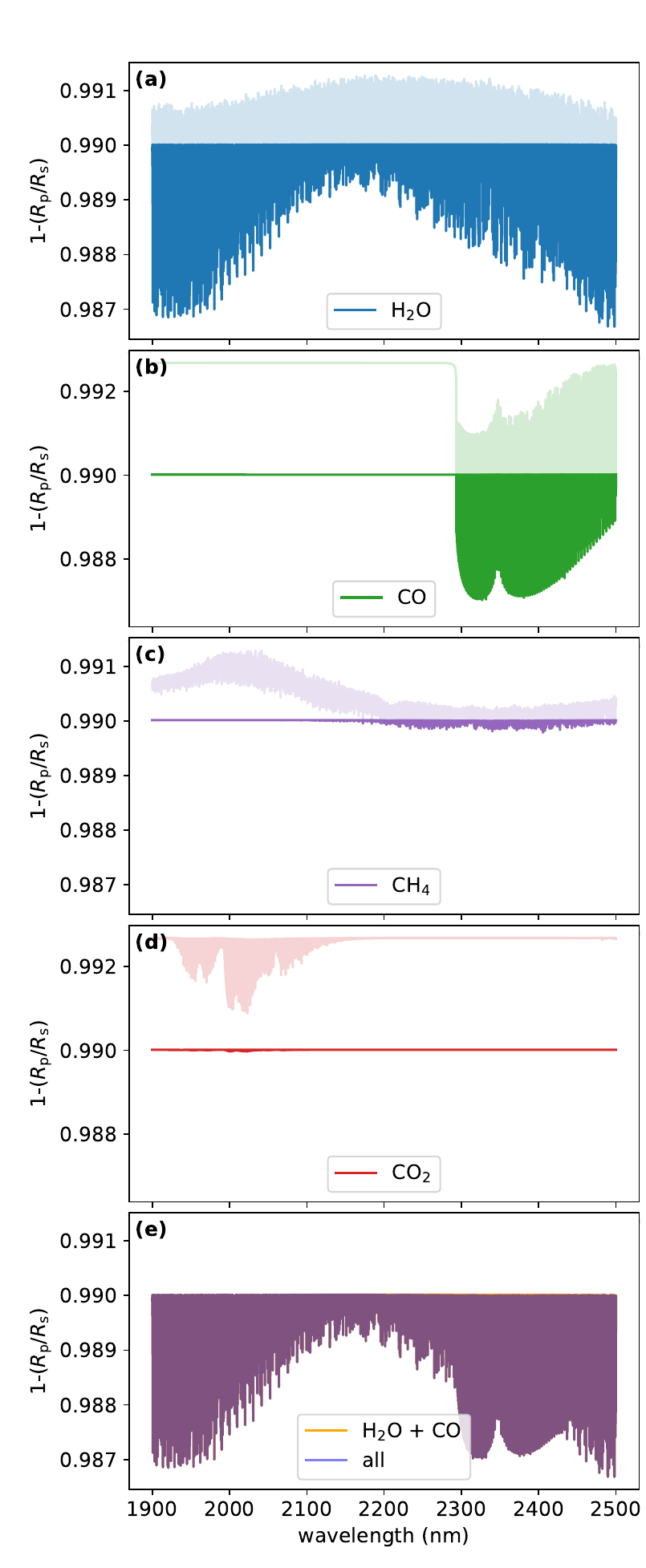}%
  \caption{Example transmission spectra for all probed species. {Panel a} shows water, {panel b} carbon monoxide, {panel c} methane, and {panel d} carbon dioxide. The lighter shaded regions indicate the features that would be visible without the cloud deck, the darker shaded regions represent the features visible despite the cloud deck. Here we use the isothermal temperature and cloud top pressure retrieved for the blueshifted signal.  {Panel e} shows a model containing all four species   plotted in light shaded blue over a model containing only H$_2$O+CO solid in orange. As both spectra are identical the resulting plot emerges as purple. Any deviation between the two models would show up as either only orange or only blue. This illustrates that methane and carbon dioxide, if present at abundances consistent with equilibrium chemistry, will not alter the spectrum with respect to a H$_2$O+CO only model. This is due to their features being hidden below the continuum, which in turn is given by the water absorption and the cloud deck.}
     \label{fig:S9}
\end{figure}
\indent In addition to the constraints on the dynamics of the planet atmosphere, we achieve tight constraints on the global C/O ratio and metallicity (C/O = $0.56_{-0.07}^{+0.05}$, [Fe/H] = $0.01_{-0.39}^{+0.36}$). These values are compatible with   the solar values and, within 2$\sigma$, with the stellar values \citep{36} (see Fig. \ref{fig:S7}). In the retrieval, the abundances for H$_2$O and CO are calculated under equilibrium chemistry assumptions based on the C/O ratio and metallicity. For the retrieved values this corresponds to mass fraction abundances of  $\log_{10}(\mathrm{H_2O}) = -2.82_{-0.39}^{+0.33}$ and $\log_{10}(\mathrm{CO}) = -2.26_{-0.39}^{+0.35}$ (volume mixing ratio abundances of $\log_{10}(\mathrm{VMR_{H2O}})= -3.71_{-0.39}^{+0.33}$ and $\log_{10}(\mathrm{VMR_{CO}}) = -3.34_{-0.39}^{+0.35})$ at the pressures above the cloud deck (see Fig. \ref{fig:S8}). In addition, the retrieved C/O ratio and metallicity are  consistent with a non-detection of CH$_4$ and CO$_2$ in the two terminators, as the features of these species would be hidden below the continuum in the presence of strong water absorption and high clouds (see \mbox{Fig. \ref{fig:S9}}).

Our retrieved temperatures agree within their uncertainties with the results of low-resolution retrievals \citep{21}. Similarly, our detection of water aligns with studies at both low and high resolution \citep{21, 36}. In the latter case, the results obtained with the SPIRou instrument yielded the detection of a single water absorption signal at high blueshift, compatible with the blueshifted peak of the water detection in this work. However, neither a significant redshifted peak nor either of the CO signals were picked up by SPIRou, even though the instrument covers a large wavelength region including the K band. With low-resolution studies unable to unambiguously identify CO in WASP-127\,b’s atmosphere \citep{21}, this had led to the assumption of a very low C/O ratio ($0.10_{-0.06}^{+0.10}$) for this planet \citep{36}, which was incompatible with our findings.  It is plausible that our successful detection of CO and the double-peaked velocity distribution is owed to the higher spectral resolution of the CRIRES$^+$ data and larger instrument throughput at the reddest end of the K band where the CO absorption lines are located. Given that each of the two peaks represents only a part of the planet spectrum, their amplitudes are smaller than the signal of a homogeneous atmosphere at \mbox{0 km s$^{-1}$} velocity would be. This effect may have misled the interpretation of weak signals of other species in this planet in the past \citep{23, 24}. Revaluation of these high-resolution data in the light of our results may affect their interpretation and consequent conclusions about the strength of absorption signals in the visible spectral range. Similarly, the consideration of a complex velocity distribution may prove rewarding in the interpretation of data from other planets that could exhibit circulation patterns similar to that of \mbox{WASP-127\,b}.

Finally, we find that the error scaling factor converges to \mbox{$\beta  = 0.6089 \pm 0.0004$}, which is of the same order of magnitude as the error scaling factors found in previous studies based on CRIRES$^+$ data \citep{29, 32}. This  suggests that the uncertainties provided by the employed version of the CRIRES$^+$ pipeline (version 1.1.4) appear to be systematically overestimated by approximately the inverse of this factor.

\section{Conclusions}\label{sec:conclusions}
We studied the transmission spectrum of the hot Jupiter \mbox{WASP-127\,b} by using high-resolution spectrophotometric observations obtained with  CRIRES$^+$ in the K band. Using the cross-correlation technique and a Bayesian retrieval framework employing a 2D modelling approach we obtained the following insights on the planet:
\begin{itemize}
\item We detected the presence of H$_2$O and CO;  the latter represents the first conclusive detection of CO in this planet.
\item We retrieved the C/O ratio and the metallicity of the planet atmosphere, which we found consistent with solar values and  with the corresponding values for the star. These results differ significantly from previous studies, which did not detect the CO signal. Our results imply that an exotic formation history for this planet is no longer required to explain this planet's chemistry.
\item We detected a double-peaked velocity profile of the atmospheric material with peaks at  approximately $\pm 8$ km s$^{-1}$, without any signal located at zero velocity, which can be explained by the presence of an equatorial jet and muted signals at the poles. 
We retrieved the velocity of the atmospheric material at the equator to be supersonic (9 km s$^{-1}$), and thus six times faster than the motion expected from tidally locked rotation, and constrained the latitudinal extent of the super-rotating jet. 
\item Finally, we retrieved the temperatures and cloud top pressures of the evening and morning terminators and of the polar regions independently. We found tentative evidence that the morning terminator is colder than the evening terminator, and that the poles are significantly colder than the evening terminator. 
\end{itemize}
Three-dimensional models of hot Jupiter exoplanets have predicted the presence of equatorial jets as well as latitudinal and longitudinal heterogeneity in temperatures, chemical compositions, and cloud conditions. Here we retrieved the atmospheric properties from the two planetary terminators whose signals close to the equator are fully resolved in velocity space, and derived constraints for the atmospheric properties of the regions at higher latitudes. Observational constraints are needed to guide theoretical work. However,  latitudinal variations have been especially difficult to constrain from observations in the past.

We showed that in the case of fast-rotating planet atmospheres, utilizing the shape of the velocity kernel of the transmission spectrum during retrieval, different regions of the planet atmosphere can be mapped. This observational input can be used to independently validate theoretical GCMs and help advance our understanding of exoplanet atmosphere circulation.

%\newpage%%%%%%%%%%%%%%%%%%%%%%%%%%%%%%%%%%%%%%%%%%
%~
%\newpage%%%%%%%%%%%%%%%%%%%%%%%%%%%%%%%%%%%%%%%%%%

%~
%\newpage%%%%%%%%%%%%%%%%%%%%%%%%%%%%%%%%%%%%%%%%%%

\begin{acknowledgements}
We thank the anonymous referee for their very helpful comments. This project is based on observations collected at the European Southern Observatory under ESO programme 108.22PH.005. CRIRES$^+$ is an ESO upgrade project carried out by Thüringer Landessternwarte Tautenburg, Georg-August Universität Göttingen, and Uppsala University. The project is funded by the Federal Ministry of Education and Research (Germany) through Grants 05A11MG3, 05A14MG4, 05A17MG2 and the Knut and Alice Wallenberg Foundation. L.N. and F.L. acknowledge the support by the Deutsche Forschungsgemeinschaft (DFG, German Research Foundation) – Project number 314665159. D.C. is supported by the LMU-Munich Fraunhofer-Schwarzschild Fellowship and by the Deutsche Forschungsgemeinschaft (DFG, German Research Foundation) under Germany´s Excellence Strategy – EXC 2094 – 390783311.
A.D.R., L.B.-Ch., and N.P. acknowledge support by the Knut and Alice Wallenberg Foundation (grant 2018.0192). O.K. acknowledges support by the Swedish Research Council (grant agreements no. 2019-03548 and 2023-03667), the Swedish National Space Agency, and the Royal Swedish Academy of Sciences. M.R. and S.C. acknowledge the support by the DFG priority program SPP 1992 “Exploring the Diversity of Extrasolar Planets” (DFG PR 36 24602/41 and CZ 222/5-1, respectively). D.S. acknowledges funding from project PID2021-126365NB-C21(MCI/AEI/FEDER, UE) and financial support from the grant CEX2021-001131-S funded by MCIN/AEI/ 10.13039/501100011033.
%more??
\end{acknowledgements}

% WARNING
%-------------------------------------------------------------------
% Please note that we have included the references to the file aa.dem in
% order to compile it, but we ask you to:
%
% - use BibTeX with the regular commands:
%   \bibliographystyle{aa} % style aa.bst
%   \bibliography{Yourfile} % your references Yourfile.bib
%
% - join the .bib files when you upload your source files
%-------------------------------------------------------------------
\bibliographystyle{aa} % style aa.bst
\bibliography{bibw127crires.bib}
\newpage
\begin{appendix}
\section{Details on data pre-processing steps}
\subsection{Extraction of the 1D spectra and auxiliary information}\label{App:extraction}
During the observation the light of the star was dispersed over the three CRIRES$^+$ detectors. This caused each of the six echelle orders (23-28), covered by the K2148 setting, to be split into three wavelength segments, yielding a total of 18 segments. Echelle order 29 is only covered for a part of the full slit (only for nodding position A) and was consequently excluded from the analysis.
To extract the spectra from the raw FITS frames we used the CRIRES$^+$ DRS pipeline \texttt{CR2RES} (version 1.1.4). In this process we arranged the time series into pairs of consecutive A and B spectra, resulting in 49 pairs. We then ran the pipeline on each pair. A and B exposures were subtracted from each other to remove the sky background and detector artifacts. The pipeline subsequently located the spectral orders and extracted the spectra using optimal extraction, accounting for the variable tilt and curvature of the slit. The wavelength solution was determined from the wavelength calibration frames taken with a Uranium-Neon lamp and a Fabry-Perot etalon. The wavelength solution provided by the pipeline is expressed in vacuum and in telluric rest frame. 
We extracted the timestamp for the starting time of each exposure from the header of the original spectrum source FITS file where it is saved as the modified Julian date (MJD) under the header-keyword ‘MJD-OBS’. We added half the exposure time and 240000.5 days to determine the mid-exposure timestamp in Julian date referenced to the coordinated universal timescale (JD$_\mathrm{UTC}$). Using these JD$_\mathrm{UTC}$ timestamps and the right ascension (RA) and declination (DEC) coordinates of the star, we calculated the time stamps in BJD$_\mathrm{TDB}$ using the \texttt{UTC2BJD} code by \citep{37} and the barycentric velocity correction $v_\mathrm{bary}$ using \texttt{barycorr.pro} \citep{38}.  The barycentric velocity correction $v_\mathrm{bary}$ was required when shifting all the spectra from telluric rest frame into the stellar and planetary rest frames.
\begin{figure}
\centering
\includegraphics[width=\hsize]{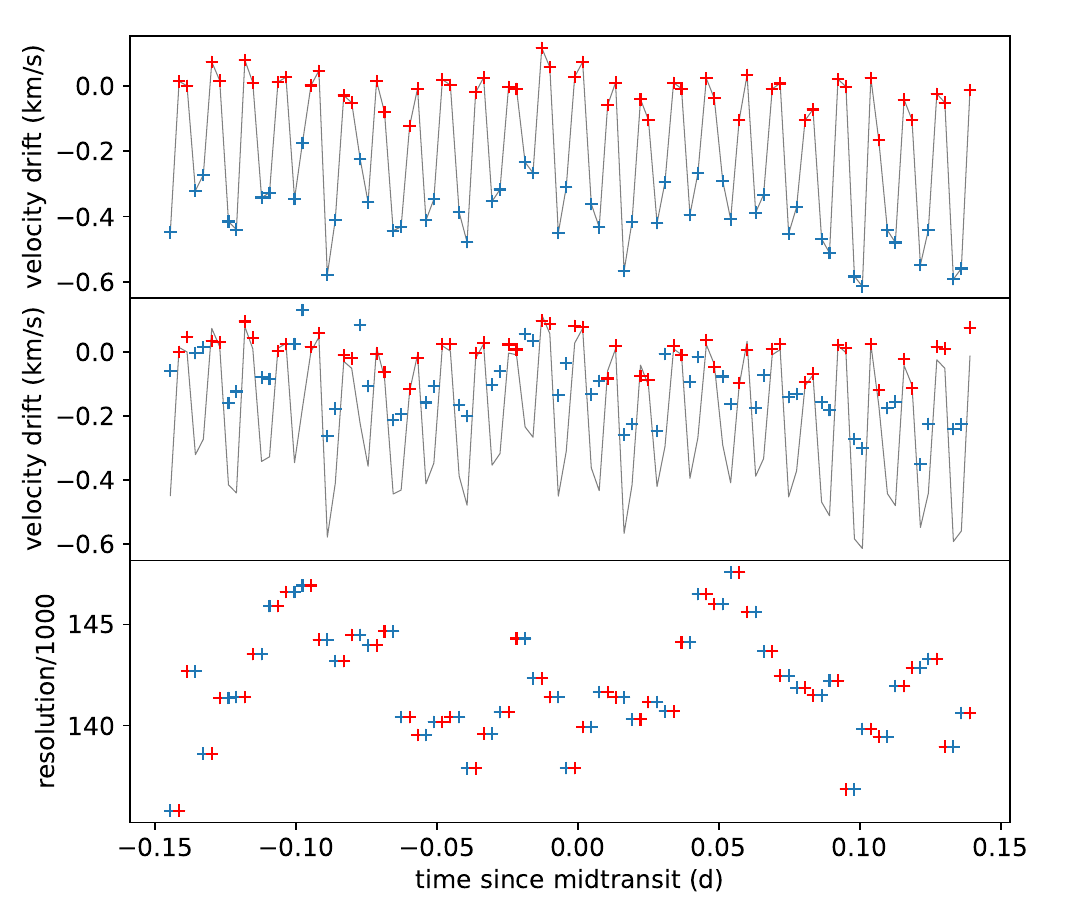}
    \caption{Effects of super-resolution on the resolution and wavelength solution of the observations. {Top panel}: results from cross-correlation of each spectrum to a reference spectrum (third spectrum of the night, nodding position B) when using the pipeline provided wavelength solution. Position A spectra are plotted in blue, position B spectra in red. {Middle panel}: same as the top panel but with the wavelength solution derived with \texttt{molecfit} from fitting of telluric lines, for one reference A and one reference B spectrum (third and fourth spectrum of the night). The previous offset is indicated by the black straight line, which is the same as the one shown in panel A. The 0.4 km s$^{-1}$ offset between A and B has been alleviated, but for both positions a drift over time remains. {Bottom panel}: mean resolution for each nodding pair from the spatial PSF measurements in each wavelength segment.}
     \label{fig:S2}
\end{figure}
\subsection{Determination of spectral resolution and correction of the wavelength solution}\label{App:wlsolution}
We found that during the night the FWHM of the spatial stellar point spread function (PSF) was $\sim$2 pixel wide. The FWHM is calculated and written in the fits headers by the pipeline for each nodding pattern. Under the premise that the width of the PSF in spatial and dispersion direction is comparable, it follows that the stellar PSF was smaller than the slit width of $0.2$''. This lead to a higher resolution than would have been achieved through slit-limited spectroscopy (also referred to as super-resolution). For comparison a PSF FWHM of 3.5 pixel would correspond to a homogeneous slit illumination and resolution of $\mathcal{R}\sim$100\,000 \citep{13}. We used the measured values for the spatial PSF for each wavelength segment and every nodding pair to calculate the respective mean spectral resolution for each nodding pair. The results are given in Fig. \ref{fig:S2} bottom panel, and show that the resolution for this night was relatively stable and scattered around a median value of $\mathcal{R}=140\,000$.

This increased wavelength resolution comes with two potential side effects:\
The first issue is that if the seeing and/or the performance of the adaptive optics changes over the night, the resolution in each spectrum might vary. However, based on the progression of the resolution over the duration of the night shown in Fig. \ref{fig:S2}, bottom panel), this was not the case for our observations. 
The second potential problem is that since the wavelength solution provided by the pipeline is obtained from observations of a calibration lamp, which homogeneously illuminates the slit, the wavelength solution is only accurate if the stellar PSF is perfectly centred within the slit. If the stellar PSF is smaller than the slit width, this is no longer guaranteed. The offset resulting from a non-centred position of the star will manifest as a wavelength shift and can be different for the A and B positions, as well as for every single spectrum, given that the position of the star may drift over the night.
We cross-correlated every A and B spectrum with a reference spectrum (the third spectrum of the night, obtained in position B) in wavelength space to check for such an effect. In this calculation, the individual wavelength solutions provided by the pipeline for A and B spectra were taken into consideration. The input spectra had been continuum-normalised as described in Sect. \ref{app:normalization} before this cross-correlation. Owing to the significantly larger number of telluric lines than stellar lines present in the spectra, the cross-correlation result was dominated by the telluric lines, finding the solution that aligns the spectra in the telluric rest frame. The results of the cross-correlation revealed that the spectra obtained in A and B positions are systematically offset from each other by approximately 0.4 km s$^{-1}$ (see Fig. \ref{fig:S2}, top panel). Moreover, we detected a small drift in the positions from the first to the last spectra of the night.

As we cannot say with certainty that the pipeline-provided wavelength solution in either the A or B position is correct (i.e. that the star was centred for these exposures), we instead obtained an empirical wavelength solution for two reference spectra, one in each position, by employing the ESO tool \texttt{molecfit} \citep{14, 39}. \texttt{Molecfit} can fit telluric lines and allows for a correction of the input wavelength solution, providing the improved solution as output information.

We repeated the cross-correlation of the spectra, using the respective improved A and B wavelength solution for all A and B spectra, and found that the systematic offset between the spectra in these two positions had disappeared, but the drift over time remained (see Fig. \ref{fig:S2}, middle panel). We finally used the \texttt{molecfit}-derived wavelength solutions for A and B, along with the velocity drift measured in the second cross-correlation as an additional time dependent correction term and interpolated all spectra to the same wavelength grid.
\subsection{\texttt{Molecfit} correction of the A and B 
reference spectra}\label{App:molecfitccf}
We performed a telluric correction of the third and fourth spectra of the night using the ESO tool \texttt{molecfit} \citep{14, 39}. \texttt{Molecfit} allows  a synthetic model of the telluric transmission to be fit to the spectra observed by ground-based observatories, and is frequently used to correct for the telluric absorption. The code is versatile and offers many settings. In our case, the fit specifically included free parameters for a polynomial model for the spectral continuum and the column depths of the molecular species water, methane, carbon dioxide, and nitrous oxide. Furthermore, we fitted for a second-order polynomial correction of the wavelength scale, which is the primary objective of our modeling.

Our CRIRES$^+$ spectra are separated into six spectral orders, each spread across three detectors. While we treated the orders separately in our modeling, we merged the spectra of the three detectors prior to the fit. As we were primarily interested in a universal adaption of the wavelength scale, we focused on the third and fourth spectra of the night (i.e. the second BA nodding pair). We found this to be more suitable than using the first pair, which exhibits lower than average signal to noise and, consistently, also the widest observed spatial PSF profile. In this way, we determined an improved wavelength scale, taking advantage of the accurately known telluric transmission spectrum, for the spectra obtained in nodding positions A and B.

\subsection{Alignment of the spectra prior to \texttt{SYSREM}}
\label{app:alignmentofspectra}
To extract the planetary signal from the observations, the stellar and telluric signals need to be removed. In this work, we employed the detrending algorithm \texttt{SYSREM} \citep{15} to facilitate the removal. \texttt{SYSREM} takes into account the uncertainties of the data but otherwise functions analogous to a principal component analysis, recognizing and removing signals that are present in all spectra. The algorithm benefits from prior continuum normalisation of the spectra and requires all spectra to be well aligned 
%in alignment (??) 
regarding the lines that we intend to remove. Unfortunately, the stellar lines and telluric lines can never both be perfectly aligned for all spectra, as the barycentric velocity changes over the duration of the observations, and the stellar lines shift due to the star’s orbit around the exo-system barycentre. In this observation these effects cause a small ($\approx$1 km s$^{-1}$) shift of the telluric lines with respect to the stellar lines over the duration of the 6.6h observation. Such a misalignment is usually ignored, as both strategies (i.e. either moving into stellar or into telluric rest frame) result in a reasonable correction of both stellar and telluric signals. Here, we repeated the alignment and normalisation for the two different cases. In the first case we moved the spectra into the telluric rest frame, so that the telluric lines, which in our case are mostly water (H$_2$O) lines, remain static in wavelength space over the observation. This should yield the best conditions for removing contamination of spurious H$_2$O signals with \texttt{SYSREM} and facilitate the detection of exoplanetary H$_2$O. In the second case, we brought all spectra into the stellar rest frame. This should yield optimal conditions for removing contaminating signals present in the star but not strongly present in the Earth’s atmosphere, such as carbon monoxide (CO), facilitating reliable detections of this molecule.

\begin{figure*}
\centering
\includegraphics[width=0.98\hsize]{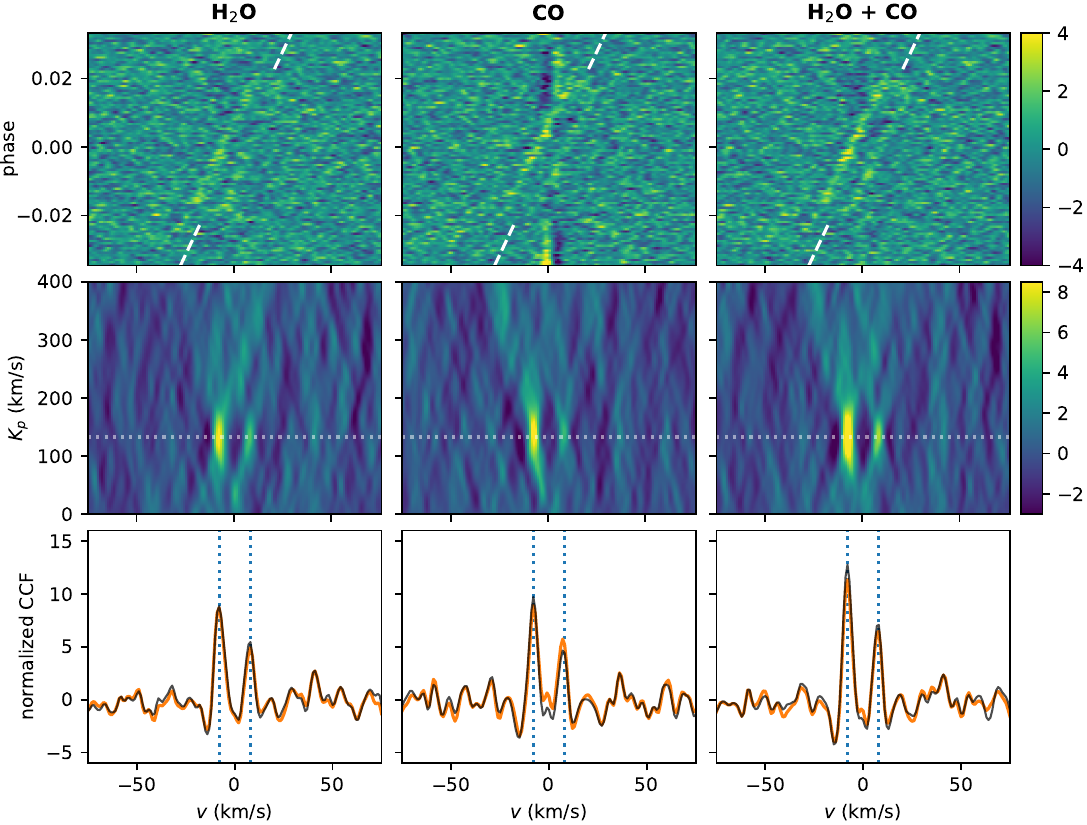}
  \caption{Results of the cross-correlation analysis for the data corrected by using \texttt{SYSREM} in telluric rest frame. Results are shown for cross-correlation with a pure H$_2$O model ({left column}), a pure CO model ({middle column}) and for a model containing both atmospheric species ({right column}) in the same manner as in as Fig. \ref{fig:1}, but with the analysis done in telluric rest frame instead of stellar rest frame.  {Top panel}: cross-correlation function for each orbital phase. For the CO case, the residual noise from stellar lines can be seen in the form of vertical structures. The radial velocity of the planetary orbital motion is indicated with a white dashed line. {Middle panel:} $K_\mathrm{p}$-$v$ map, i.e. the CCF co-added using different values for $K_\mathrm{p}$. The white dotted line indicates the value for $K_\mathrm{p}$ found in the literature. {Bottom panel:} cut through the $K_\mathrm{p}$-$v$ map at the literature $K_\mathrm{p}$ value (black solid lines). The curves resulting from the analysis performed in stellar rest frame, which are shown in Fig. \ref{fig:1}, are overplotted (orange solid lines) to facilitate comparison between the CCFs resulting from the two different approaches.}
     \label{fig:S4}
\end{figure*}

In this work, the cross-correlation analysis was performed for both versions of the data:  aligned to stellar rest frame before \texttt{SYSREM} (results shown in Fig. \ref{fig:1}) and aligned to telluric rest frame before \texttt{SYSREM} (results shown in Fig. \ref{fig:S4}). These results show that for the 'telluric rest frame' analysis significant stellar residuals remain in the data. In comparison, detrending in stellar rest frame improved the removal of these stellar signals without compromising the removal of telluric residuals. 
 To align the spectra, we applied the \texttt{molecfit}-derived wavelength solutions and the cross-correlation derived velocity drift of the solution over the night. We then interpolated every spectrum to the grid of the reference spectrum in nodding position B (third spectrum of the night). In the case of aligning to stellar rest frame, we additionally shifted by the velocity drift of the stellar lines over the night, assuming a circular orbit. The stellar velocity was calculated using 
\begin{equation}
v_\mathrm{star}(t) = - K_\mathrm{s} \sin\left(2 \pi \xi\left(t\right)\right) - v_\mathrm{bary}(t)+ v_\mathrm{system}
,\end{equation}
where $\xi$ is the phase of the orbit, $K_\mathrm{s}$ is the stellar orbital velocity semi-amplitude, $v_\mathrm{bary}$ is the barycentric velocity shift, and $v_\mathrm{system}$ is the constant velocity of the observed system with respect to the Solar System barycentre. We calculated the phase of the orbit based on the ephemeris information from \citep[][listed in Table \ref{tab:seidelparameters}]{24}.

\subsection{Normalisation of the spectra}\label{app:normalization}
We normalised the spectra as follows: The spectra are read into a data array of the dimensions [98 $\times$ 2048 $\times$ 18], corresponding to number of spectra in the time series, number of wavelength channels per segment, and number of wavelength segments, respectively. The same process is done for the uncertainties of the data. We then performed a continuum normalisation for each of the 18 wavelength segments independently using the following protocol: We computed a master spectrum as the mean of all spectra. We identified the positions of wavelength regions affected by telluric lines from a theoretical telluric spectrum, calculated with \texttt{molecfit} using the same wavelength sampling as our data. In this \texttt{molecfit} spectrum, we identified all points lower than 96\% of the continuum as telluric lines and flagged their position in the data analysis of the science spectra. Additionally, we flagged any ‘not a number’ (NaN) values in the data. We iteratively fitted a second-order polynomial function to the master stellar spectrum in which both telluric lines and NaNs had been masked. We identified and iteratively masked stellar lines by rejecting points that had values significantly below the polynomial fit continuum and then repeating the fit. This is performed over five iterations, with the rejection criterium going from less than 80\% of the continuum in the first iteration, to less than 90\% in the second, and then to less than 95\% in subsequent iterations. We then divided all spectra by the master spectrum and fitted the residuals (while masking telluric lines, NaNs, and the identified stellar lines) with another second-order polynomial function. We iterated this fitting step for three iterations, and in each step, we rejected any points deviating by more than 20\% of the fit as bad pixels. The final continuum correction of each individual spectrum was, thus, the multiplication of the polynomial fit to the continuum of the master stellar spectrum and the polynomial fit to the residuals of the individual spectrum divided by the master spectrum.
We divided all spectra by their continuum fits. We subsequently checked column by column how many pixels we had identified as either NaNs or bad pixels. If more than 3 pixels in one column (i.e. more than three exposures) were affected, we masked the entire column. If 3 or fewer pixels were affected, we replaced these bad points by linearly interpolated values.
\subsection{Objective determination of the optimal number of \texttt{SYSREM} iterations}\label{sec:bestit}
\begin{figure*}
%\centering
\includegraphics[width=\hsize]{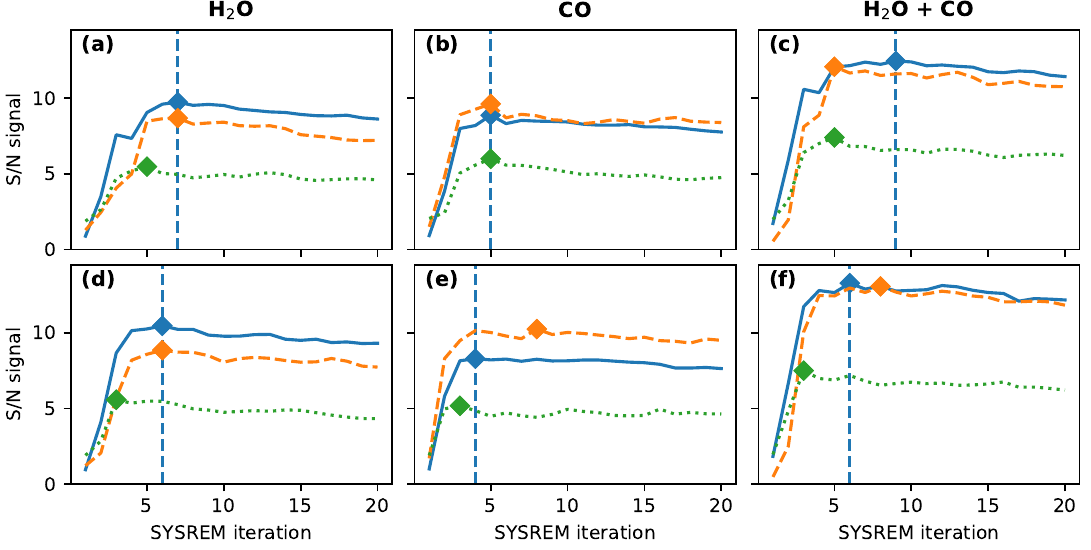}
  \caption{Determination of the best \texttt{SYSREM} iteration. 
Shown is the progression of the S/N of the recovered signal with the number of \texttt{SYSREM} iterations for the injected signal (blue solid line), the blueshifted signal (orange dashed line), and the redshifted signal (green dotted line). {Panels a, b and c} show the values for \texttt{SYSREM} performed in (quasi-)stellar rest frame, leading to the results shown in Fig. \ref{fig:1}. {Panels d, e and f} show the values for \texttt{SYSREM} performed in telluric rest frame leading to the results shown in Fig. \ref{fig:S4}. In all cases the diamonds indicate the peak signal position.}
     \label{fig:S3}
\end{figure*}

%\subsection{Optimal number of \texttt{SYSREM} iterations}\label{sec:bestit}
We determine the ‘best’ iteration by injecting a synthetic model into the data. The model, which contained both H$_2$O and CO and a cloud layer at $\log_{10}(P_\mathrm{c} /(bar)) = -3$, was injected into the data at planet orbital velocity before the normalisation step. We repeat the entire analysis for the dataset with the injected signal in the same manner as for the original data. We then normalise the $K_\mathrm{p}$-$v$ map with the injected signal using the noise value calculated for the $K_\mathrm{p}$-$v$ map without the injected signal, and then subtract the results from the analysis without injected signal from the analysis with injected signal. This effectively isolates the injected signal from any small background signals present in the original data. This procedure is in essence the same as was described by \cite{18}, with the execution only slightly differing as they subtract the original data from the injected data before calculating the $K_\mathrm{p}$-$v$ map. In Fig. \ref{fig:S3}, we plot the progression of the injected signal strength with \texttt{SYSREM} iterations, and for comparison, also plot the progression of the two real (not-injected) signals. The development of the S/N with increasing \texttt{SYSREM} iteration shows that after the fourth iteration, the S/N does not improve drastically anymore. Finally, we choose the iteration at which the injected signal is recovered at the strongest S/N for the H$_2$O+CO case (iteration 9) as the input for the atmospheric retrieval. \FloatBarrier
\FloatBarrier
~~~
\FloatBarrier
\newpage
~~~
\FloatBarrier
\newpage
\onecolumn
\section{Posterior distributions}
~~\\
~~

\FloatBarrier
%\vspace{1.5cm}
\begin{figure}[h]
\centering
\includegraphics[width=0.90\hsize]{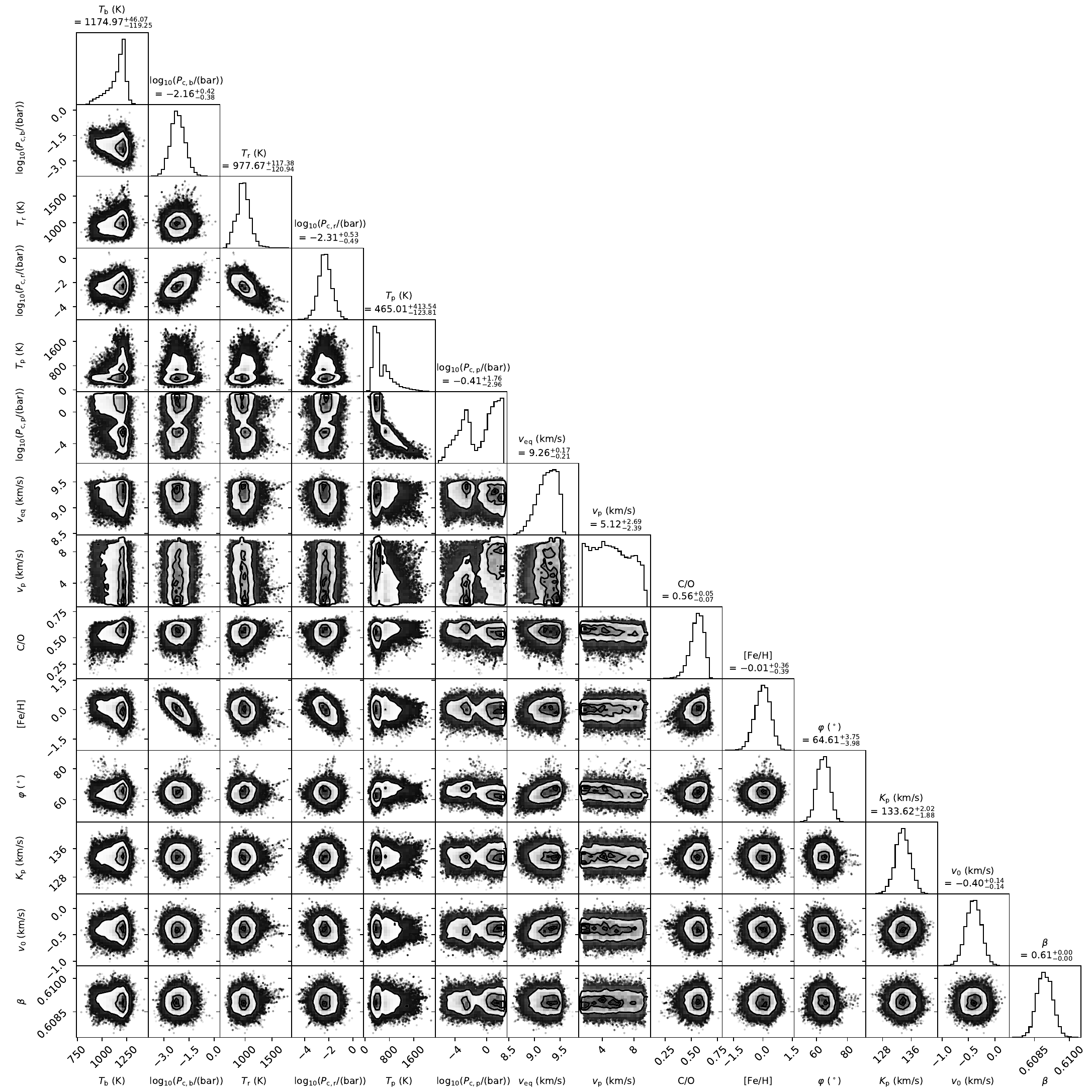}
  \caption{Results of the retrieval. The first six distributions represent the isothermal temperatures $T$ and cloud top pressures $\log_{10}(P_\mathrm{c}$) of the three independently modelled atmospheric regions (index b stands for the blueshifted signal, r for the redshifted signal and p for the polar regions). These values are followed by the distributions for the maximal velocity at the equatorial regions ($v_\mathrm{eq}$) and the one for the polar regions ($v_\mathrm{p}$). The last six distributions are showing the shared atmospheric parameters C/O ratio and metallicity [Fe/H], the latitudinal angle that separates the equatorial jet from the poles $\varphi$, the system parameters $K_\mathrm{p}$ and $v_0$, as well as a noise scaling factor $\beta$. }
     \label{fig:S6}
\end{figure}
\FloatBarrier
%\clearpage
\vspace*{1.5cm}
\begin{figure}[h]
\includegraphics[width=0.90\hsize]{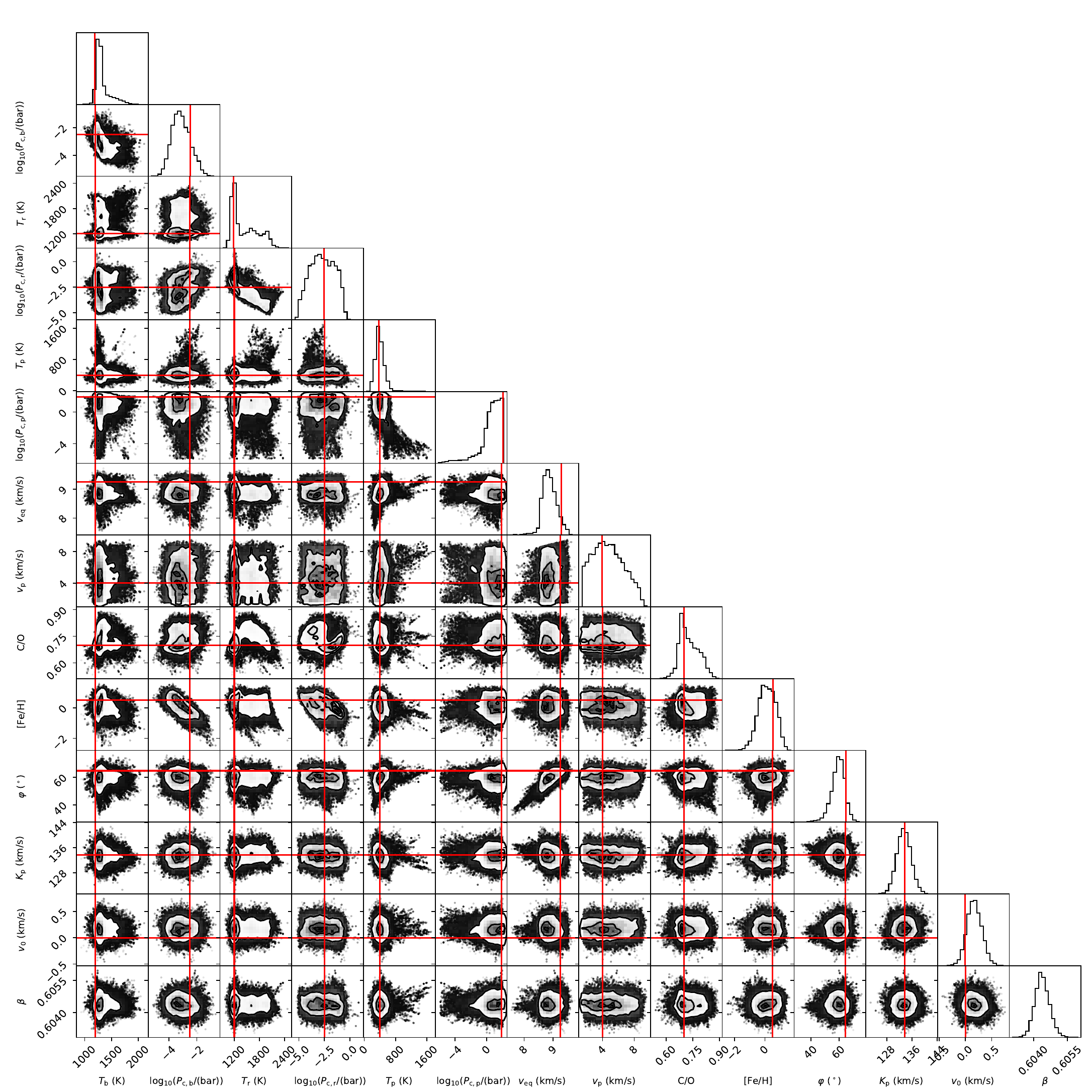}
  \caption{Results of the injection recovery test. The parameters are shown in the same order as in Fig. \ref{fig:S6}. In red we plot the location of the injected parameters. No value was injected for $\beta$, as the noise scaling factor is determined from the data.}
     \label{fig:injectionrecoverycorner}
\end{figure}
~
\end{appendix}
\end{document}